\documentclass[preprint2]{aastex61}

\PassOptionsToPackage{breaklinks=true}{hyperref}
\def\chandra    {\emph{Chandra}}

\received{2025 March 5}
\revised{2025 July 9}
\accepted{2025 July 11 - ApJ, in press}

\shorttitle{The structure of the giant radio fossil in Ophiuchus}
\shortauthors{Giacintucci et al.}

\begin{document}

\title{The structure of the giant radio fossil in the Ophiuchus galaxy cluster}

\correspondingauthor{Simona Giacintucci}
\email{simona.giacintucci.civ@us.navy.mil}

\author{S. Giacintucci}
\affiliation{Naval Research Laboratory, 
4555 Overlook Avenue SW, Code 7213, 
Washington, DC 20375, USA}
\author{M. Markevitch}
\affiliation{NASA/Goddard Space Flight Center,
Greenbelt, MD 20771, USA}
\author{T. Clarke}
\affiliation{Naval Research Laboratory, 
4555 Overlook Avenue SW, Code 7213, 
Washington, DC 20375, USA}
\author{Daniel R. Wik}
\affiliation{Department of Physics \& Astronomy, The University of Utah, 115 South 1400 East, Salt Lake City, UT 84112, USA}

\begin{abstract}

We present high-sensitivity follow-up observations of the giant fossil radio lobe in the Ophiuchus galaxy cluster with the upgraded Giant Metrewave Radio Telescope (uGMRT) in the 125--250 MHz and 300--500 MHz frequency bands. The new data have sufficient angular resolution to exclude compact sources and enable us to trace the faint extended emission from the relic lobe to a remarkable distance of 820 kpc from the cluster center. The new images reveal intricate spatial structure within the fossil lobe, including narrow (5-10 kpc), long (70-100 kpc) radio filaments embedded within the diffuse emission at the bottom of the lobe. The filaments exhibit a very steep spectrum ($S_\nu\propto \nu^{-\alpha}$ with $\alpha \sim 3$), significantly steeper than the ambient synchrotron emission from the lobe ($\alpha \sim 1.5-2$); they mostly disappear in recently-published MeerKAT images at 1.28 GHz. Their origin is unclear; similar features observed in some other radio lobes typically have a spectrum flatter than that of their ambient medium. These radio filaments may trace regions where the magnetic field has been stretched and amplified by gas circulation within the rising bubble. The spectrum of the brightest region of the radio lobe exhibits a spectral break, which corresponds to a radiative cooling age of the fossil lobe of approximately 174 Myr, giving a date for this most powerful AGN explosion.

\end{abstract}



\section{Introduction}

Galaxy clusters --- the most massive gravitationally bound systems in the Universe --- are permeated by a hot, X-ray emitting plasma known as the intracluster medium (ICM). This plasma, with temperatures reaching $10^{7-8}$K ($kT\sim1-10$ keV), contains several times more mass than the cluster galaxies and is thus the dominant baryonic component of clusters. 

In most of the cluster volume, the radiative cooling time of the ICM exceeds the age of the cluster. However, in the cores of clusters with sharply peaked density profiles, often marked by giant cD galaxies \citep[e.g.,][]{1984ApJ...276...38J,1998MNRAS.298..416P}, the plasma temperature within the central $r\sim100$ kpc region ($\sim5\%$ of the cluster virial radius) exhibits a steep decline towards the core \citep[e.g.,][]{2005ApJ...628..655V}. Simultaneously, the gas density sharply increases, resulting in a highly X-ray luminous core, where the radiative cooling time is shorter than the cluster's age.
This core is thermally unstable: as the plasma cools, its density increases to maintain pressure equilibrium, which further accelerates the cooling process. 
However, the observed mass of ``warm'' gas (with temperatures just below $kT\sim 1$ keV) in clusters with the highest cooling rates is only a few percent of the amount predicted by this simple physical picture \citep[e.g.,][]{2006PhR...427....1P}.
Since X-ray cooling is directly observed by X-ray telescopes, a compensatory heating mechanism is necessary to prevent such instability. The leading candidate is radio-mode feedback from 
the active galactic nucleus (AGN) hosted by the central galaxy of the cluster \citep[see e.g.,][for a review]{2022hxga.book....5H}.
In this scenario, the AGN jets, fueled by the accretion of the infalling cooling gas, inflate buoyant bubbles within the hot X-ray emitting gas. This process should inject enough energy back into the system to compensate for cooling \citep[e.g.,][]{2012ARA&A..50..455F,2012NJPh...14e5023M,2015Natur.519..203V,2017MNRAS.466..677G}.

However, the precise mechanism through which AGNs transfer energy to the ICM within cool cores remains poorly understood. It is unclear whether this energy transfer always occurs gradually or ``gently". Evidence suggests that in some systems, AGN heating can be violent, involving large-scale shocks. Notable examples include the clusters Hydra A \citep{2005ApJ...628..629N,2007ApJ...659.1153W}
and MS\,0735.6+7421 \citep{2005Natur.433...45M}.
Moreover, violent AGN outbursts, such as Hydra A and MS\,0735.6+7421, appear to have deposited the majority of their enormous mechanical energy ($10^{60-61}$ erg) far beyond the central cooling region (i.e., at $\sim 150-200$ kpc from the center). This raises a critical question: how do these extraordinary energetic events fit into the AGN feedback scenario, which is crucial for stabilizing cluster cores? 
It is the dense central gas that poses the most significant challenge in terms of cooling and stability.

The Ophiuchus cluster presents a possible dramatic counterexample to the gentle AGN feedback scenario. {\em Chandra}\/ X-ray observations revealed a prominent concave surface brightness edge just outside the cool core, $r\sim 120$ kpc from the center \citep{2016MNRAS.460.2752W}, which appears to be the inner wall of an enormous X-ray cavity, spanning approximately $400-500$ kpc in diameter. Using low-frequency (72--240 MHz) radio observations by the Murchison Widefield Array (MWA) GLEAM survey\footnote{Galactic and Extragalactic All-sky MWA survey \citep{2017MNRAS.464.1146H}.}
and the Giant Metrewave Radio Telescope (GMRT), \citet[][hereafter G20]{2020ApJ...891....1G} discovered a giant, ultra steep-spectrum ($\alpha=2.4$\footnote{The radio spectral index $\alpha$ is defined according to $S_{\nu} \propto \nu^{-\alpha}$, where $S_{\nu}$ is the flux density at the frequency $\nu$.}) diffuse source that fits inside this X-ray super-cavity. The brightest part of this source, the one that closely follows the the X-ray wall, has also been detected in recent MeerKAT images at 1.28 GHz \citep{botteon25}. 

In G20, the radio data were compared with the X-ray images from \chandra\ and \emph{XMM-Newton} (the latter image, but adding more X-ray data, is also shown here in Fig.\ 5b). A diffuse radio minihalo entirely fills the Ophiuchus bright cool core \citep[G20; see also][]{2009A&A...499..371G, 2010A&A...514A..76M,botteon25}. The X-ray data do not show any merging subclusters at position of the ultra steep-spectrum diffuse source that might suggest that it is another minihalo. The steep-spectrum source is also clearly spatially separated from the central minihalo, so it does not appear to be a "minihalo within a giant radio halo" scenario \cite[e.g.,][]{venturi17, 2024A&A...692A..12V}.

G20 posited that the radio source and the X-ray cavity is a fossil radio lobe resulting from an exceptionally powerful outburst from the AGN in the Ophiuchus center. The energy required to displace the gas and create this cavity is estimated to be few $\times 10^{61}$ erg \citep{2016MNRAS.460.2752W}, making it the most powerful AGN outburst observed in any galaxy cluster. A difficulty for this scenario is the apparent lack of a counterpart lobe on the opposite side of the core. G20 proposed that the outburst has happened so long ago that the other lobe either faded out of the radio band and/or was swept and disrupted by gas sloshing. Our new radio data adds an interesting alternative, and we will return to this question in Section \ref{sec:disc}.

Recent {\em XMM-Newton} X-ray observations of the region revealed a candidate weak shock front in the ICM on the outer side of the cavity, likely driven by the expansion of the radio lobe as it inflated the giant cavity (Markevitch et al.\ 2025, in preparation).
Similarly to other systems with large cavities accompanied by shocks, such as Hydra A and MS\,0735.6+7421, this detection indicates a strong impact of the AGN outburst on the ICM, raising the question of how the cool core may have survived such an episode. 

This paper presents a deep radio follow-up of the fossil radio lobe in Ophiuchus using the upgraded GMRT (uGMRT) in Bands 2 (125--250 MHz) and 3 (300--500 MHz). 
These new observations enable us to map the faintest radio emission within the cavity, investigate the spectral properties of the lobe, and estimate the age of the lobe based on its radio spectrum.

We use a $\Lambda$CDM cosmology with H$_0$=70 km s$^{-1}$ Mpc$^{-1}$,
$\Omega_m=0.3$ and $\Omega_{\Lambda}=0.7$. At the redshift of 
Ophiuchus ($z=0.028$), $1^{\prime\prime}$ = 0.562 kpc. All errors are quoted at the 68\% confidence level.

\begin{deluxetable*}{ccccccccc}
\tablecaption{uGMRT observations}
\label{tab:radioobs}
\tablehead{
\colhead{Project} &  \colhead{Frequency range} & \colhead{Bandwidth}& \colhead{Channel} & \colhead{Observation} & \colhead{Total time} &  \colhead{Time on target} & Flux density\\
\colhead{Code}    & \colhead{(MHz)}     & \colhead{(MHz)}    & \colhead{number} & \colhead{Date}    & \colhead{(Hour)} & \colhead{(Hour)} & calibrator \\
}
\startdata
34$_{-}$070 & 125-250 & 200   & 8192 &2018 May 07 & 9 & 7.5 & 3C 48\\ 
34$_{-}$070 & 250-500 & 200   & 8192 & 2018 May 08  & 9 & 7.5  & 3C 48\\
\enddata \tablecomments{Column 1: project code. Columns 2--4: observing
  frequency range, bandwidth, and number of spectral channels. Column 5: observation date.
  Column 6: total observation time, including calibration overheads.  Column 7: time on target. Column 8: flux density calibrator.}
\end{deluxetable*}


\begin{deluxetable}{ccc}
\tablecaption{uGMRT band sub-divisions}
\label{tab:subbands}
\tablehead{
\colhead{Sub-band} &  \colhead{Central frequency} & \colhead{Bandwidth} \\
\colhead{} &  \colhead{(MHz)} & \colhead{(MHz)} \\
}
\startdata
B2--01 & 147 & 61 \\
B2--02 & 186 & 16 \\
B2--03 & 202 & 16 \\
B2--04 & 219 & 16 \\
B2--05 & 235 & 16 \\
\phantom{0}B2--06$^{a}$ & 251 & 16 \\
\phantom{0}B2--07$^{a}$ & 267 & 16 \\
B3--01 & 313   & 26 \\
B3--02 & 340.5 & 29 \\
B3--03 & 371   & 32 \\
B3--04 & 404   & 34 \\
B3--05 & 440   & 38 \\
B3--06 & 479.5 & 21 \\
\enddata \tablecomments{
$^{a}$ Failed calibrationbecause heavily affected by radio frequency interference.}
\end{deluxetable}

\section{Radio observations}
\label{sec:radio}

We observed the Ophiuchus cluster with the uGMRT in Bands 2 (125-250 MHz) and 3 (300-500 MHz) during Cycle 34, on May 7th and 8th, 2018. Each band was observed for a total of 9 hours, resulting in 7.5 hours of on-source observation time after accounting for calibration overheads. All data were collected in spectral-line mode using the wide-band back-end with a bandwidth of 200 MHz, 8192 frequency channels, and an integration time of 4 seconds. Table \ref{tab:radioobs} summarizes the observational details.

Data calibration was performed using the automated Source Peeling and Atmospheric Modeling (SPAM) pipeline \citep{2009A&A...501.1185I,2017A&A...598A..78I}, specifically modified for uGMRT data\footnote{http://www.intema.nl/doku.php?id=huibintema
spampipeline}. The pipeline automatically divides the wide observing bandwidth (200 MHz) into narrower sub-bands for independent processing. 
For Band 2, the data were split into one sub-band centered at 147 MHz (61 MHz bandwidth) and six additional sub-bands ranging from 178 MHz to 275 MHz, each 16 MHz wide (details in Table 2). Band 3 followed a similar approach with six sub-bands spanning 300 MHz to 500 MHz (Table 2). Each sub-band underwent individual processing in SPAM, adopting a standard calibration scheme. This scheme includes bandpass and complex gain calibration, followed by self-calibration procedures (both direction-independent and direction-dependent). The flux density scale was set using the calibrator source 3C\,48 and \cite{2012MNRAS.423L..30S}. Phase 
calibration was carried using the default sky model in SPAM, which is based on the NVSS\footnote{NRAO VLA Sky Survey, \cite{1998AJ....115.1693C}}, WENSS\footnote{Westerbork Northern Sky Survey, \cite{1997A&AS..124..259R}}, VLSSr\footnote{Very Large Array Low-frequency Sky Survey Redux, \cite{2014MNRAS.440..327L}}, SUMSS\footnote{Sydney University Molonglo Sky Survey, \cite{2003MNRAS.342.1117M}}, and MGPS-2\footnote{Molonglo Galactic Plane Survey 2nd Epoch, \cite{2007MNRAS.382..382M}} catalogs. Sub-bands B2-05 and B2-06  (Tab.~2) were excluded due to significant radio frequency interference encountered during the observation.

After self-calibration, the FITS visibility data were converted into measurement sets using the Common Astronomy Software Applications (CASA\footnote{https://casa.nrao.edu.}, version 5.6) package. The data were then imaged using WSClean with the joint-channel and multi-scale deconvolution options
\citep{2014MNRAS.444..606O,2017MNRAS.471..301O}. Images were produced at the central frequencies listed in Table 2: 147 MHz (B2-01), 210 MHz (B2-02 to B2-05), and 400 MHz (B03-01 to B3-06). During imaging, we employed various weighting schemes, including uniform and Briggs "robust" weights from -0.5 to +0.5 \citep{1995PhDT.......238B}, and applied tapers to the $uv$-data. This resulted in images with angular resolutions ranging from 6$^{\prime\prime}$ ($\sim 3$ kpc) to 120$^{\prime\prime}$ ($\sim$67 kpc). Lower resolution imaging was not performed at 147 MHz due to limitations in data quality.

Image details are summarized in Table 3. All images were corrected for the uGMRT primary beam response using the task PBCOR in the Astronomical Image Processing System \citep[AIPS\footnote{http://www.aips.nrao.edu.},][]{2003ASSL..285..109G} package. For Band 3 images, we used primary beam shape parameters from a December 2018 GMRT memo\footnote{http://www.ncra.tifr.res.in/ncra/gmrt/gmrt-users/
observing-help/ugmrt-primary-beam-shape}. For the Band 2 images, we used parameters from a June 2022 memo\footnote{www.gmrt.ncra.tifr.res.in/doc/Beam$_{-}$shape$_{-}$
band$_{-}$2.pdf}. Systematic amplitude uncertainty is estimated to be within 10$\%$ in Band 2 and 5$\%$ in Band 3 \citep{2022ApJ...934..170L}. 

\begin{deluxetable*}{ccccccccc}
\tablecaption{uGMRT images}
\label{tab:images}
\tablehead{
\colhead{\#} & \colhead{Frequency} & \colhead{Sub-band} & \colhead{FWHM}, {\em p.a.} & \colhead{rms} & \colhead{Weighting} & \colhead{$uv$ taper} & \colhead{{\em uv} range } & Figure \\
\colhead{} & \colhead{(MHz)}     &  \colhead{} & \colhead{($^{\prime \prime} \times^{\prime \prime}$, $^{\circ}$)} &  \colhead{(mJy beam$^{-1}$)} & \colhead{} & \colhead{($^{\prime\prime}$)} & \colhead{(k$\lambda$)} & \\
}
\startdata
1 & 147 & B2--01 & 24$\times$17, 2 &  1.2  & Robust 0 & $-$ & 0.02--14 & \ref{fig:fila2}(a) \\
2 & 147 & B2--01 & 24$\times$24, 0 &  1.8  & Robust 0 & $-$ & 0.04--14 &  \ref{fig:ridge} \\
3 & 164 & B2--01 to 03 & 24$\times$24, 0 & 0.9 & Robust 0 & $-$ & 0.04--14 & \ref{fig:spix1}(a) \\
4 & 194 & B2--02 \& 03  & 24$\times$24, 0&  0.8  & Robust 0 & $-$ & 0.04--14 & \ref{fig:ridge} \\
5 & 210 & B2--02 to 05 & 16$\times$9, 18 & 0.3  & Robust $-$0.5 & $-$ & 0.03-21 & \ref{fig:fila1}\\
6 & 210 & B2--02 to 05 & 60$\times$60, 0     & 1.3 & Robust 0 & 33 & 0.04-21 & \ref{fig:low}(a) \\
7 & \phantom{0}210$^{\star}$ & B2--02 to 05 & 60$\times$60, 0     & 1.3 & Robust 0 & 33 & 0.04-21 & \ref{fig:spix1}(c) \\
8 & 210 & B2--02 to 05 & 120$\times$120, 0   & 2.2 & Robust 0 & 60 & 0.04-21 & \ref{fig:low}(c) \\ 
9 & 227 & B2--04 \& 05 & 24$\times$24, 0 &  0.4  & Robust 0 & $-$ & 0.04--14 & \ref{fig:ridge} \\
10 & 333 & B3--01 \& 02 & 24$\times$24, 0 &  0.05  & Robust 0 & $-$ & 0.04--14 & \ref{fig:ridge}  \\
11 & 400 & B3--01 to 06 & 7$\times$4, $-1$ & 0.03  & Uniform & $-$ &  0.04-42 & \ref{fig:400ext}, \ref{fig:tails} \\
12 & 400 & B3--01 to 06 & 9$\times$6, 12  & 0.024 & Robust 0 & $-$ & 0.04-42 &  \ref{fig:fila2}(b), \ref{fig:field}\\
13 & 400 & B3--01 to 06 & 18$\times$18, 0 & 0.025 & Robust 0 & $-$ & 0.04-21 & \ref{fig:rgb} \\
14 & 400 & B3--03 \& 04 & 24$\times$24, 0 &  0.05  & Robust 0 & $-$ & 0.04--14 &  \ref{fig:ridge}\\
15 & 400 & B3--01 to 06 & 60$\times$60, 0 & 0.3 & Robust 0 & 40 & 0.04-21 & \ref{fig:low}(b) \\
16 & \phantom{0}400$^{\star}$ & B3--01 to 06 & 60$\times$60, 0 & 0.3 & Robust 0 & 40 & 0.04-21 & \ref{fig:spix1}(c) \\
17 & 400 & B3--01 to 06 & 120$\times$120, 0 & 0.9 & Robust 0 & 100 & 0.04-42 & \ref{fig:low}(d)\\
18 & \phantom{0}400$^{\star}$ & B3--01 to 06 & 120$\times$120, 0 & 0.8 & Robust 0 & 100 & 0.04-42 & \ref{fig:400ext} \\
19 & 467 & B3--05 \& 06 & 24$\times$24, 0 &  0.09  & Robust 0 & $-$ & 0.04--14 &  \ref{fig:ridge}\\
\enddata \tablecomments{Column 1: image number. Column 2: central frequency. Column 3: sub-band. Column 4: FWHM of the radio beam.  
Column 5: rms noise level ($1\sigma$). Columns 6--8: weighting, gaussian taper and $uv$ range used in the imaging. 
Column 9: figure in the paper.
$^{\star}$ image obtained after subtraction of compact sources from the $uv$ data.}
\end{deluxetable*}

\begin{figure*}
\centering \epsscale{1.1}
\includegraphics[width=15cm]{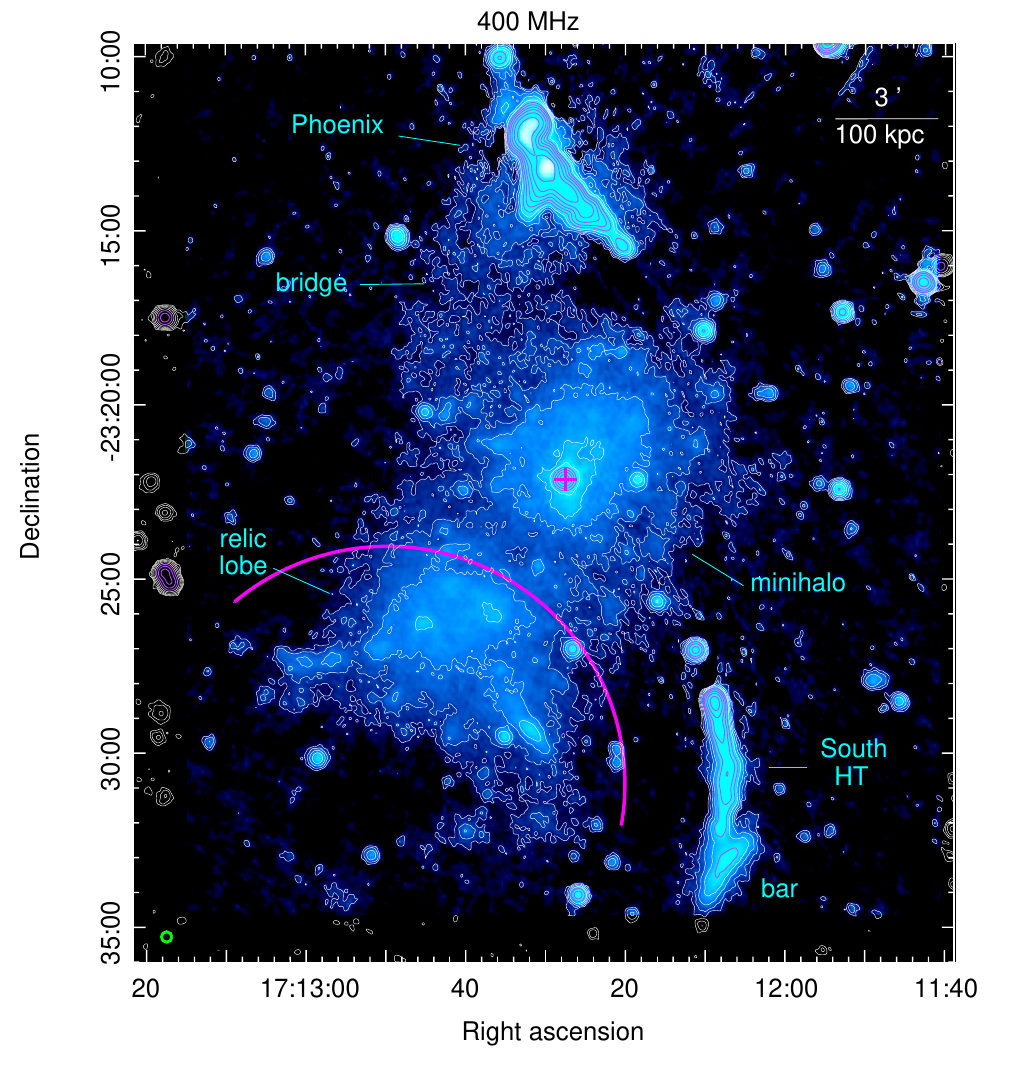}
\caption{uGMRT 400 MHz image (colors and contours) of the Ophiuchus cluster at $18^{\prime\prime}$ resolution (image  $\#13$ in Tab.~3). 
The region is $25^{\prime}\times25^{\prime}$ (0.8 Mpc$\times$0.8 Mpc) 
and the position of the BCG at the cluster center 
is marked by a magenta cross. The magenta arc is a portion of a $r=6^{\prime}.8=230$ kpc circle that traces the X-ray edge seen in the {\em Chandra} and 
{\em XMM-Newton} images \citep[][G20]{2016MNRAS.460.2752W}. The radio beam is shown as a green circle in the bottom-left corner. The rms noise is $1\sigma=25$ $\mu$Jy beam$^{-1}$. Contours are spaced by a factor 
of 2 starting from $+5\sigma$.} 
\label{fig:rgb}
\end{figure*}

\section{Radio images}
\label{sec:radioim}

This section presents new uGMRT images of the Ophiuchus cluster at 147 MHz, 210 MHz, and 400 MHz. We begin by describing the global radio emission across the cluster region (Section \ref{sec:region}) and then focus on the detailed properties of the fossil radio lobe located outside the cluster core (Section \ref{sec:lobe}). 

\begin{figure*}
\centering \epsscale{1.1}
\includegraphics[width=13cm]{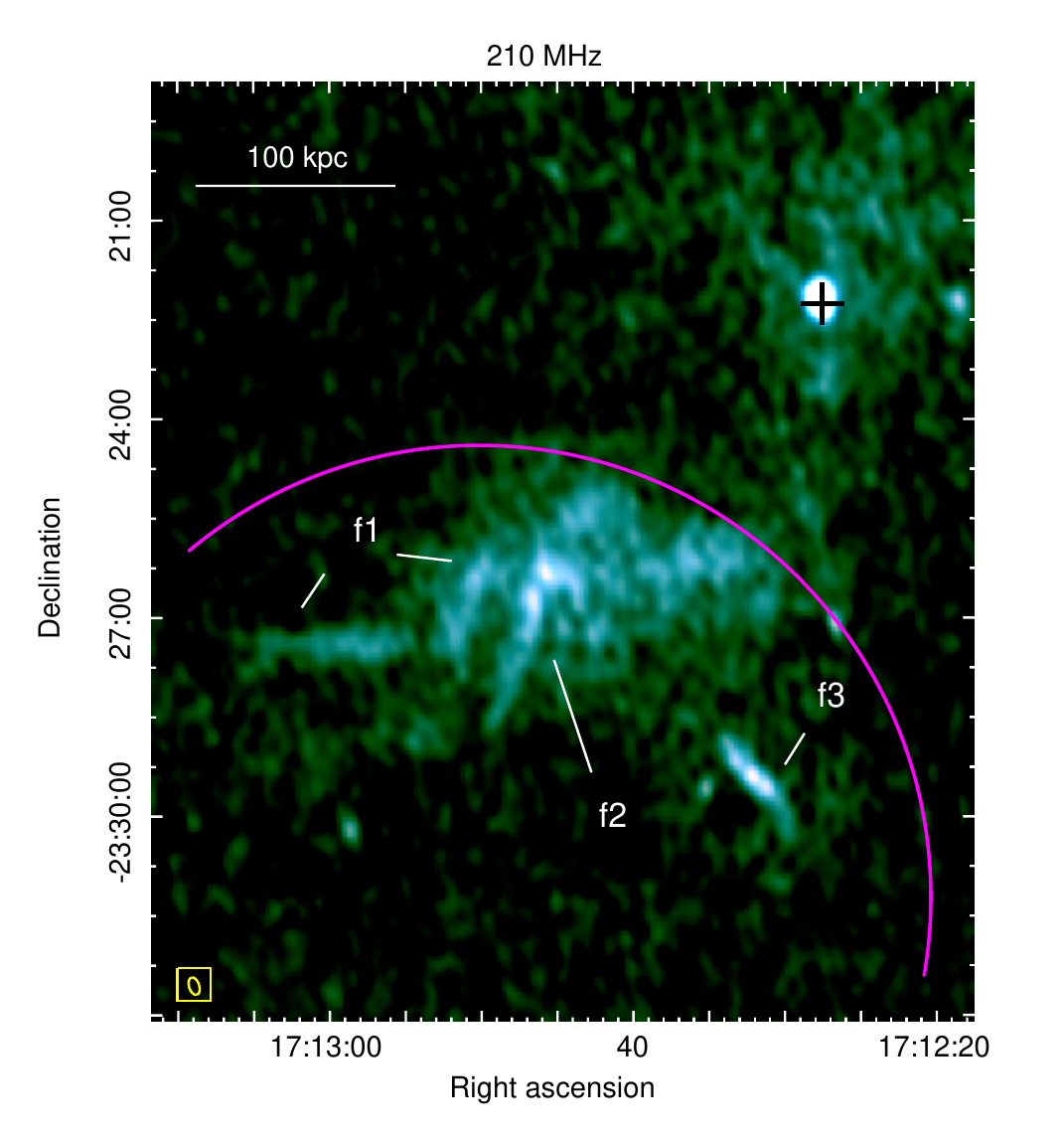}
\caption{uGMRT high-resolution image of the relic lobe at 210 MHz (image $\#5$ in Tab.~3). 
The beam size is $16^{\prime\prime}\times9^{\prime\prime}$ (yellow boxed ellipse in the bottom-left corner)
and noise level is $1\sigma=0.3$ mJy beam$^{-1}$. The black cross and magenta arc mark the 
BCG and X-ray edge, respectively. The brightest radio filaments within the lobe are labelled.} 
\label{fig:fila1}
\end{figure*}

\subsection{Radio emission in the Ophiuchus region}\label{sec:region}

Figure \ref{fig:rgb} presents our uGMRT image of the Ophiuchus cluster at 400 MHz, obtained at a resolution of $18^{\prime\prime}$. The image encompasses a region of $25^{\prime}\times25^{\prime}$, corresponding to 0.8 Mpc $\times$ 0.8 Mpc. The position of the brightest cluster galaxy (BCG) at the cluster center is indicated by a magenta cross. A magenta arc traces the concave X-ray edge observed in previous X-ray studies \citep[][G20]{2016MNRAS.460.2752W}. The relic radio lobe, discussed in detail in Section \ref{sec:lobe}, is located within this X-ray edge. An image at the native resolution of $9^{\prime\prime}\times6^{\prime\prime}$, spanning a broader $\sim 1^{\circ}\times1^{\circ}$ field centered on Ophiuchus is provided in the Appendix (Fig.~\ref{fig:field}; image $\#12$ in Tab.~3).

As visible in Figure \ref{fig:rgb}, the cluster inner region is filled with diffuse emission from a radio minihalo, extending approximately 200 kpc in radius \citep[][G20]{2009A&A...499..371G, 2010A&A...514A..76M,botteon25}. The minihalo enshrouds a compact radio source associated with the BCG (magenta cross). 

North of the core, a prominent elongated and amorphous source, labeled as "phoenix" is visible. The nature of this object remains unclear. It lacks an optical counterpart and displays a complex morphology characterized by radio-emitting threads, transverse features and vortex-like structures which are more clearly evident in the higher-resolution image  presented in Figure ~\ref{fig:tails}(a) in the Appendix \citep[image $\#11$ in Tab. 3; see also images in][]{2010A&A...514A..76M,2016MNRAS.460.2752W,botteon25}. \cite{2016MNRAS.460.2752W} proposed that this source might be a radio phoenix – fossil radio plasma injected into the ICM by past AGN activity, which has been subsequently re-energized and compressed adiabatically by the passage of a cluster shock wave \citep{2001A&A...366...26E}. 
\cite{botteon25} offered an alternative interpretation. They pointed to a striking similarity between the observed transverse radio features, or "ribs," and numerical simulations by \cite{upreti24}. These simulations indicate that such structures can arise from kink instability within a radio galaxy jet, which causes the jet to wiggle and bend. \cite{botteon25} identified a galaxy located at the south-west extremity of the radio source as a possible optical host (ID 80 in their paper). No active radio core is detected at the position of this galaxy and the bulk of the extended emission is offset from this candidate host, suggesting a scenario where the jet activity paused and then restarted.

Notably, the phoenix appears connected, in projection, to the central minihalo by a 200 kpc ``bridge'' of lower surface brightness emission ($\sim 0.3$ $\mu$Jy arcsec$^{-2}$). This bridge is also clearly detected in the new 1.28 GHz MeerKAT by \cite{botteon25} with a striking arc-shaped structure.

South-west of the cluster core, a radio galaxy with a head-tail (HT) morphology extends for approximately 220 kpc. Our uGMRT image reveals an interesting feature at the end of the tail: a 100 kpc bar misaligned by $\sim 50^{\circ}$ with respect to the tail axis. Figure \ref{fig:tails}(b) in the Appendix (image $\#$11 in Tab. 3) shows the bar resolving into two distinct parallel filaments of emission at higher resolution. Similar bar-like structures have been observed in other NATs, such as A2443 \citep{2011AJ....141..149C} and the A3558 complex \citep{2022A&A...660A..81V}, but their origin remains unclear. A comparison with numerical simulations \citep{2019ApJ...876..154N, 2019ApJ...885...80N} suggests that these features might represent toroidal, ring-like structures viewed in projection. These structures could be driven by the interaction between the radio plasma within the jets and a cluster shock wave.

In addition to the Phoenix and South HT, the Ophiuchus region hosts a number of extended radio galaxies with intriguing morphologies. Their images are presented in the Appendix (Figs. 12 and 13). For new MeerKAT L-band images of these sources, which reveal similarly intriguing features to those described here, we refer to \cite{botteon25}.

\subsection{The relic lobe}\label{sec:lobe}

We leverage the high angular resolution of our new uGMRT data to investigate the fine structure within the fossil radio lobe. To map the full extent of its emission on larger scales, we utilize images with coarser resolution.

\begin{figure*}
\centering
\includegraphics[width=17cm]{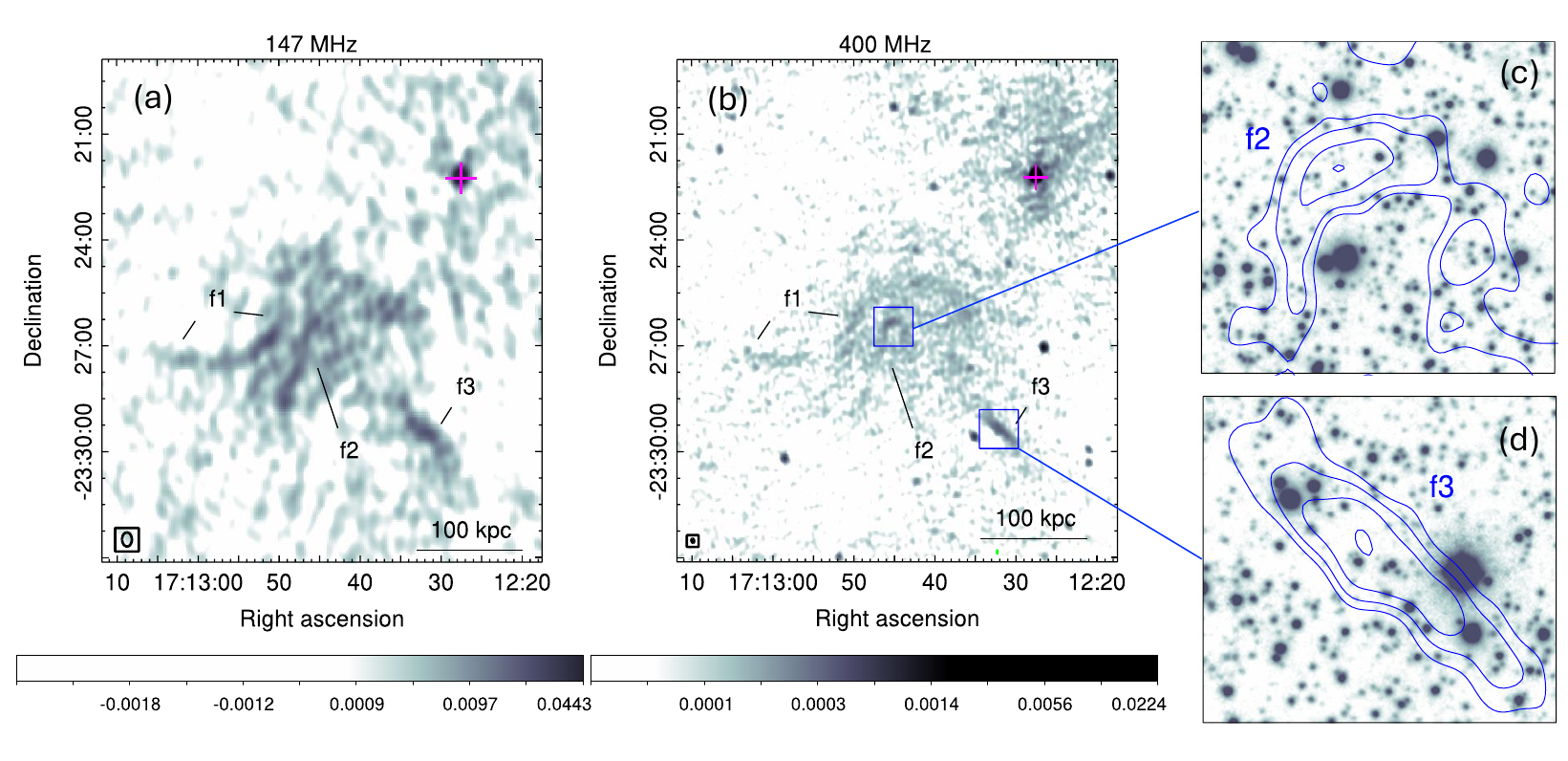}
\caption{uGMRT high-resolution images of the relic lobe at (a) 147 MHz (image $\#1$ in Tab.~3) 
and (b) 400 MHz (\#12 in Tab.~3). The beam size is $24^{\prime\prime}\times17^{\prime\prime}$ and $9^{\prime\prime}\times6^{\prime\prime}$ respectively (boxed ellipses in the bottom-left corner). The noise is $1\sigma=1.2$ and 0.024 mJy beam$^{-1}$, respectively. The magenta cross marks the BCG location. The color bar units are Jy beam$^{-1}$. 
Filaments within the relic lobe are labelled as in Fig.~\ref{fig:fila1}. Panels (c) and (d) show  the 400 MHz contours of the brightest regions of f2 and f3 (from (b)) on optical images from the DeCam Galactic plane survey. Contours scale by a factor of $\sqrt2$ from $\sim5 \sigma$.}

\label{fig:fila2}
\end{figure*}

\subsubsection{High-resolution images}\label{sec:high}

Figure~\ref{fig:fila1} presents a color image of the 210 MHz emission from the innermost part of the relic lobe, with a resolution of $16^{\prime\prime}\times9^{\prime\prime}$. This region lies just inside the X-ray edge (arc). High-resolution images of the same region at 147 MHz ($24^{\prime\prime}\times17^{\prime\prime}$) and 400 MHz ($9^{\prime\prime}\times6^{\prime\prime}$) are shown in Figure~\ref{fig:fila2}.
At all frequencies, this inner region of the lobe exhibits a complex and non-uniform structure, with bright filaments embedded within a more diffuse and fainter component. The most prominent filaments, labeled f1, f2, and f3, have lengths of $\sim$70-100 kpc and widths of $\sim$5-10 kpc (in projection). 
Given their linear morphology and extent, these structures could be tailed radio galaxies superimposed on the lobe emission. The low Galactic latitude of Ophiuchus ($b=9.3^{\circ}$) results in a high density of foreground stars in optical images, complicating the identification of potential optical hosts. Despite this challenge, a comparison of optical images from the Dark Energy Camera (DECam) Galactic plane survey \citep{2018ApJS..234...39S}
with the filaments brightest regions at our highest resolution 
(Fig.~\ref{fig:fila2}(c) and (d) for f2 and f3, respectively) reveals no associated galaxies at the radio peaks, which disfavors a radio tail interpretation. We will explore alternative possibilities for their nature in Section \ref{sec:orig_fila}.

\subsubsection{Low-resolution images}\label{sec:low}

\begin{figure*}
\gridline{\fig{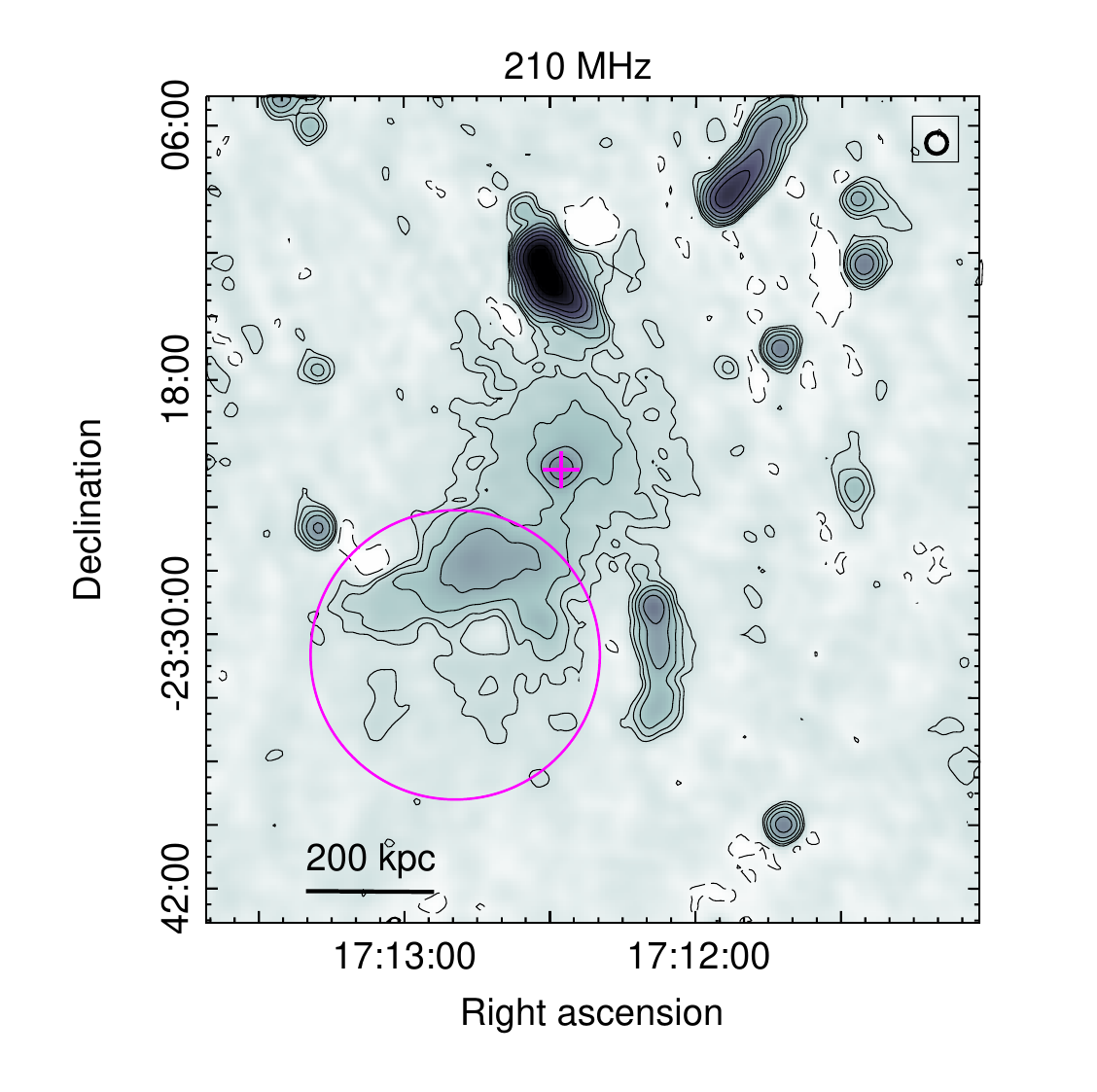}{0.5\textwidth}{(a)}
          \fig{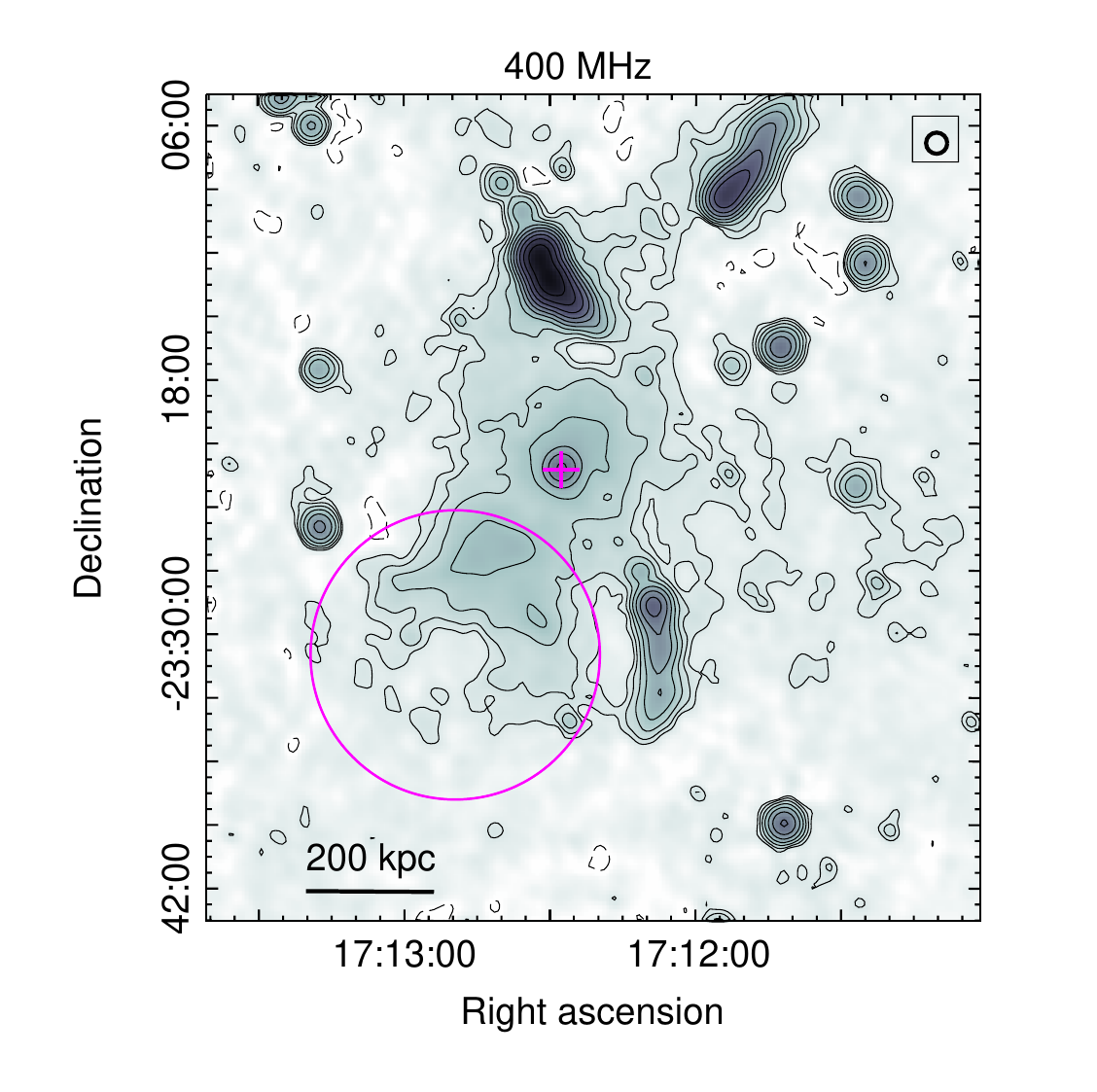}{0.5\textwidth}{(b)}}
\gridline{\fig{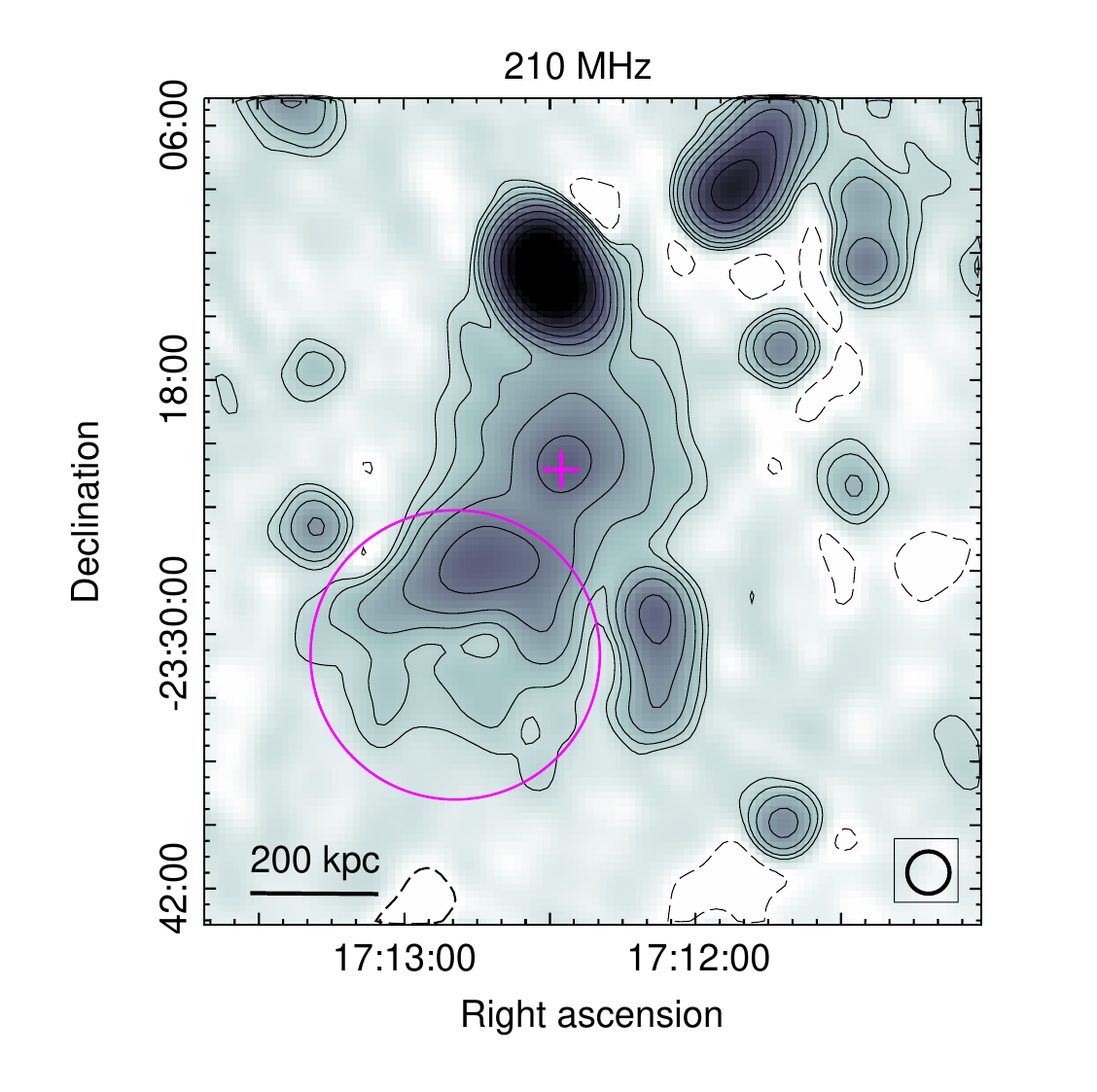}{0.5\textwidth}{(c)}
          \fig{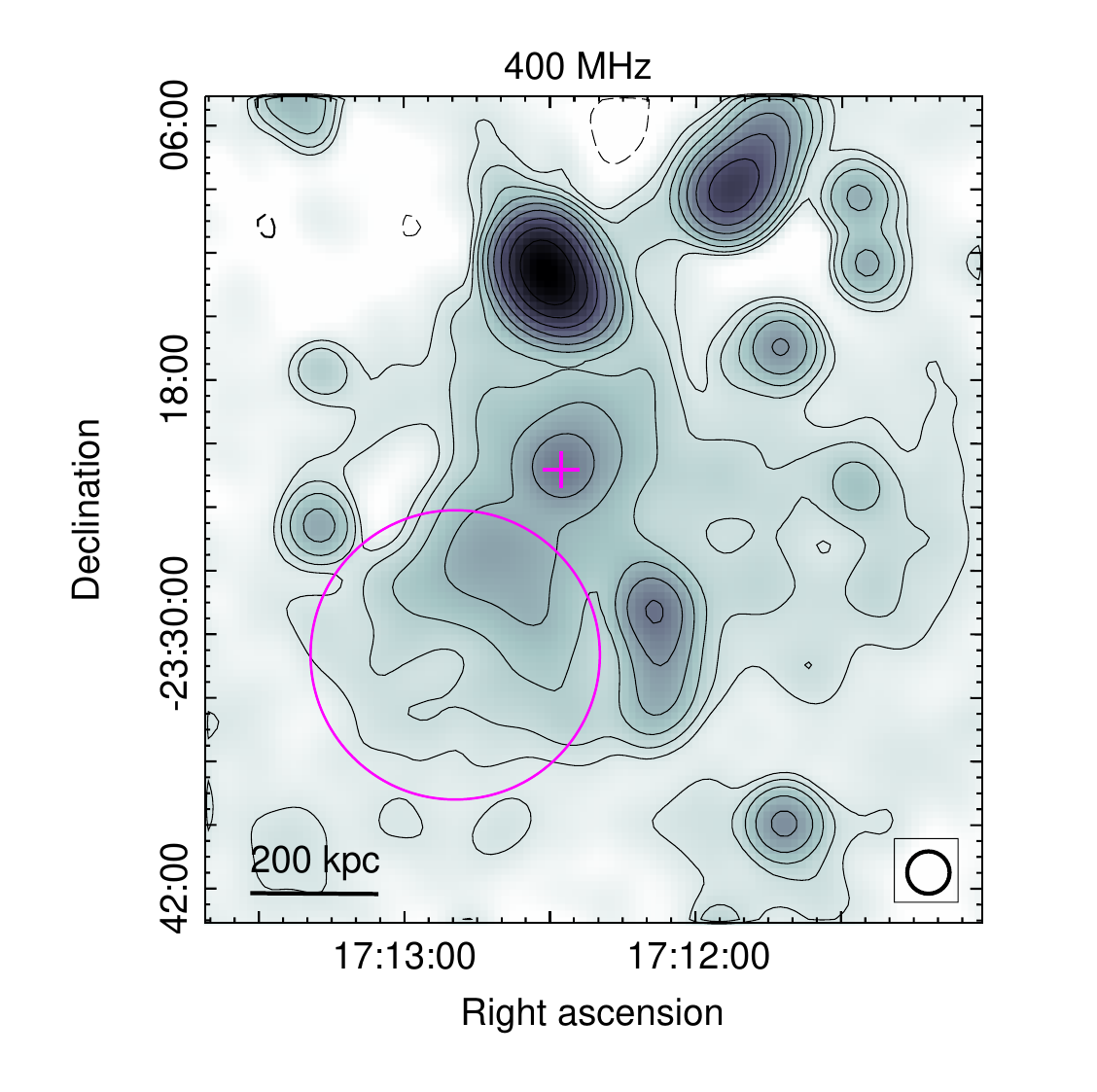}{0.5\textwidth}{(d)}}
\caption{uGMRT 210 and 400 MHz images at (a,b) $60^{\prime\prime}$ resolution (images $\#6, 15$ in Table 3) and (c,d) $120^{\prime\prime}$ resolution ($\#8, 17$ in Table 3).  In all panels, the black boxed circle shows the beam size. 
Contours scale by a factor of 2 from $\sim 3\sigma$= 4, 1, 6.6 and 3 mJy beam$^{-1}$, respectively. Dashed contours are drawn at $-3\sigma$. 
The magenta cross marks the BCG location. The magenta circle, with radius of $6.^{\prime}8$ = 230 kpc, encompasses the fossil radio lobe and at the same time traces the X-ray edge.}
\label{fig:low}
\end{figure*}

The radio filaments described above are embedded within lower surface brightness extended emission from the lobe. This extended emission is better imaged at lower angular resolution. Figure~\ref{fig:low} presents images at 210 MHz and 400 MHz with resolutions of $60^{\prime\prime}$ (a, b) and $120^{\prime\prime}$ (c, d), respectively. These images were obtained using $uv$ tapers to de-emphasize long baselines, enhancing the visibility of extended emission on larger scales (see Table 3 for details). These lower-resolution images allow us to trace the faintest and outermost emission of the relic lobe out to $\sim 400$ kpc from the X-ray edge.

\begin{figure*}
\centering 
\includegraphics[width=18cm]{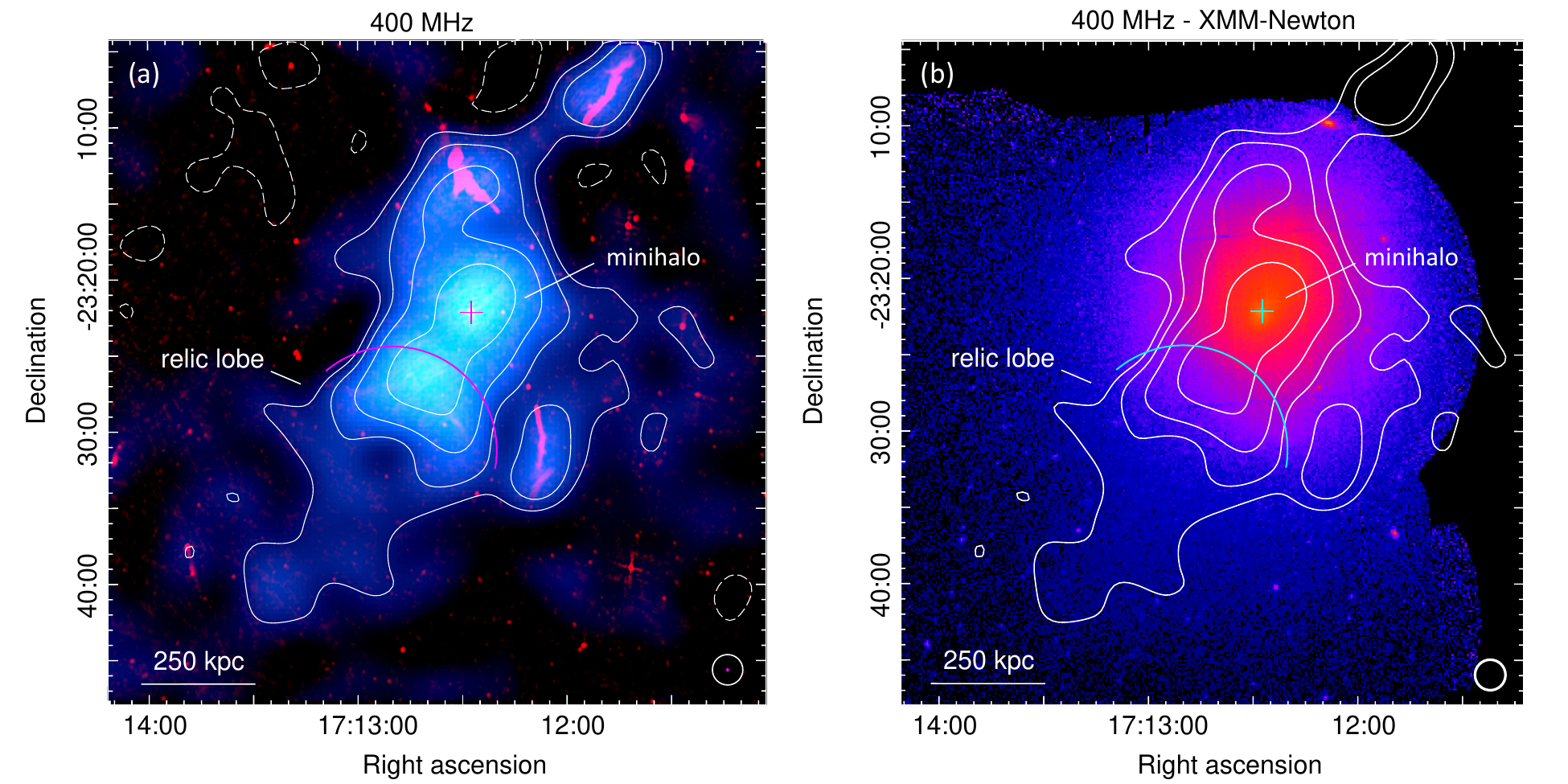}
\caption{(a) uGMRT composite image at 400 MHz. Magenta: high-resolution image
($7^{\prime\prime}\times4^{\prime\prime}$) made with uniform weights, 
including compact radio sources ($\# 11$ in Tab.~3). 
Blue and white contours: low-resolution image ($120^{\prime\prime}$) of the 
extended emission after subtraction of compact radio sources 
($\# 18$ in Tab.~3). The beam size is shown by the 
white circle and magenta circles in the bottom-right corner. 
The rms noise is $1\sigma=0.030$ mJy beam$^{-1}$ (high resolution) 
and $1\sigma=0.8$ mJy beam$^{-1}$ (low resolution). 
Contours scale by a factor of 2 from $+3\sigma$. Contours at $-3\sigma$ 
are drawn as dashed. (b): 
{\em XMM-Newton} X-ray image in the $0.4–7.2$ keV band (same as in G20 but adding newer data; Markevitch et al.\ 2025), binned to the $7.5''$ pixel size. Radio contours from panel (a) are overlaid in white. In both panels, the BCG location is marked by a cross and the arc traces the edge of the X-ray cavity, filled by the relic lobe. The cluster cool core 
contains the radio minihalo.}
\label{fig:400ext}
\end{figure*}

Figure~\ref{fig:400ext} shows a 400 MHz low-resolution image (blue and white contours) after subtracting compact radio sources. These sources were identified on an image created using uniform weights (shown in magenta) and then removed from the $uv$ data.
In the subtracted image, the relic lobe appears to extend further southeastward, reaching a maximum distance of approximately 700 kpc from the X-ray edge and a total of $\sim$ 820 kpc from the cluster center (marked by a cross). Extended emission to a similar radius was previously detected by the MWA GLEAM survey at lower frequencies (G20).

\begin{table}
\caption{Total flux density of the relic lobe}
\label{tab:flux}
\begin{center}
\begin{tabular}{lccc}
\hline\noalign{\smallskip}
\hline\noalign{\smallskip}
 & Frequency & Flux density & Reference \\
 &  (MHz)    & (Jy)       & \\
\noalign{\smallskip}
\hline\noalign{\smallskip} 
VLSSr &  74   &  $18.0\pm3.6$ & G20  \\ 
GLEAM &  88   &  $11.6\pm1.8$ & G20 \\ 
GLEAM & 119   & $5.0\pm0.7$   & G20     \\
GMRT & 153    &  $1.7\pm0.3$ & G20 \\ 
GLEAM & 155   & $2.1\pm0.5$ & G20      \\
GLEAM & 200   & $1.5\pm0.3$   & G20  \\
uGMRT & 210  & $1.2\pm0.1$ & this work \\
GMRT & 240    &   $0.82\pm0.01$ & G20   \\
uGMRT & 400 & $0.29\pm0.01$ & this work \\
VLA & 1477    & $0.012\pm0.001$ & G20  \\
\hline{\smallskip}
\end{tabular}
\end{center}
NOTE -- Measured within a $r=6^{\prime}.8$ circular region, centered on 
RA$_{\rm J2000}$=17h12m50.2s and DEC$_{\rm J2000}$=$-23$d31m13s (magenta circle in Fig.~4; see also G20). 
All flux densities have been re-scaled to the \cite{2017ApJS..230....7P} scale. 
\end{table}

\begin{figure}
\centering \epsscale{1.1}
\includegraphics[width=9cm]{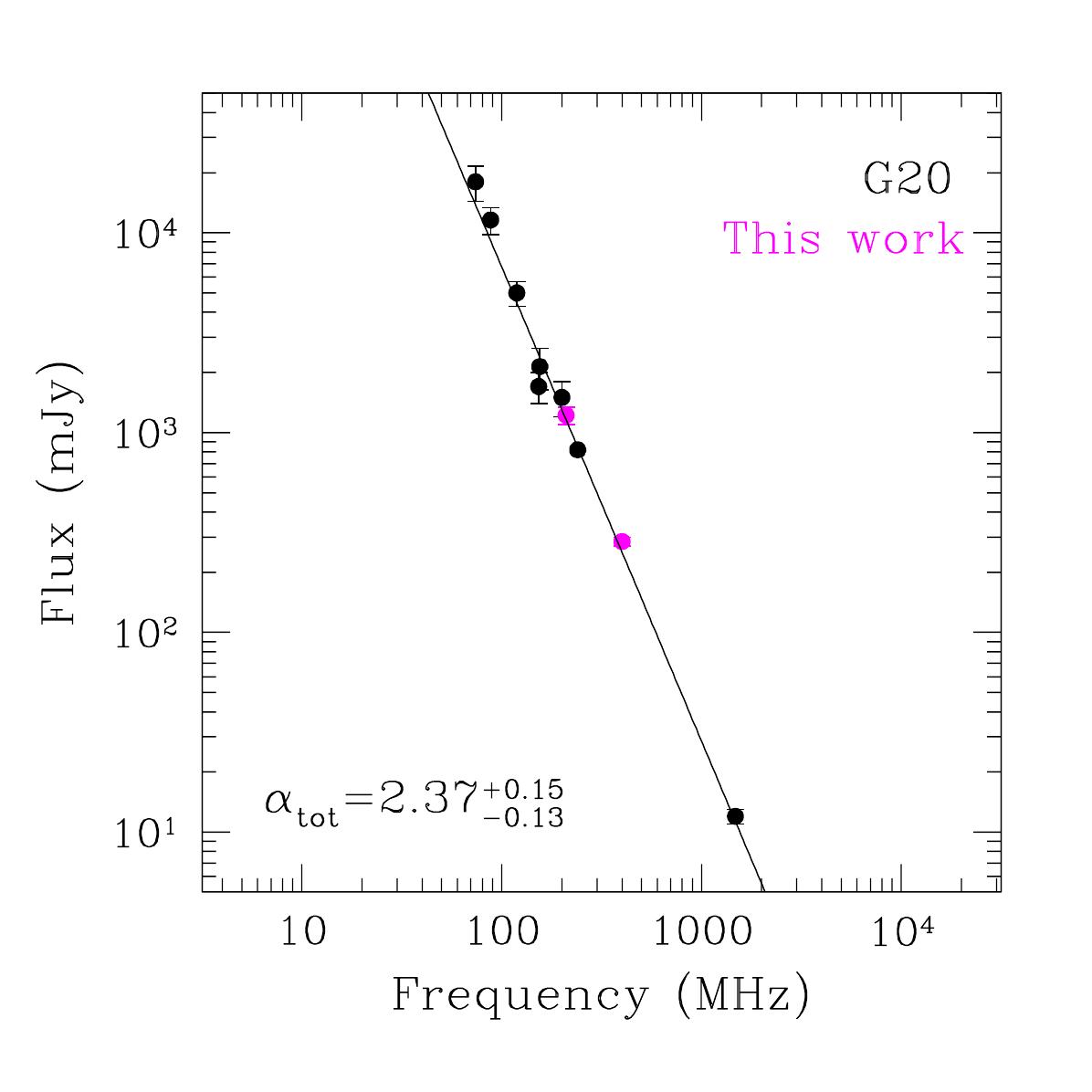}
\hspace{-0.3cm}
\caption{Integrated radio spectrum of the relic
lobe between 74 MHz and 1477 MHz using the flux densities 
in Table 4. The new uGMRT measurements 
at 210 MHz and 400 MHz are shown in magenta.
The solid line is the best-fit power-law model, whose 
slope is also reported.}
\label{fig:spectrum}
\end{figure}

\section{Integrated radio spectra}\label{sec:int}

We measured the total flux density of the relic lobe in the new uGMRT low-resolution images at 210 MHz and 400 MHz (Fig.~\ref{fig:low}) and compared these measurements to previous results from G20 to obtain an updated integrated spectrum (Section~\ref{sec:sp_tot}). 
Due to significant contamination from radio-frequency interference at short baselines and increased noise levels, the 147 MHz images only provide adequate mapping of the innermost region of the relic lobe at high resolution (Fig.~3a). The data quality is insufficient for a reliable low-resolution mapping of the entire extended lobe emission, and therefore we excluded this frequency from our total flux density measurements. 

We utilized our high-resolution images at all frequencies to derive the radio spectrum of the innermost and brightest region of the lobe, including the embedded filaments (Section~\ref{sec:sp_inner}).

\subsection{Total emission}\label{sec:sp_tot}

To measure the total flux density of the relic lobe in Fig.~\ref{fig:low} (c,d), we used the same circular region with a radius of $6^{\prime}.8$, centered at RA$_{J2000}$=17h12m50s, DEC$_{J2000}$=$-$23d31m13s, as employed in G20. This region approximates the size of the X-ray cavity inferred from the curvature radius of the concave X-ray edge (circle in Fig.~\ref{fig:low}).

Table 4 summarizes the lobe flux densities at all frequencies, including the G20 measurements. Errors encompass both image noise and calibration uncertainties and are estimated as 

\begin{equation}
\Delta S_{\nu} = \sqrt{(f \times S_{\nu})^2 + N_{\rm beam}(\sigma_{\rm rms})^2}
\end{equation}

where $f$ is the absolute flux density calibration uncertainty, $S_{\nu}$ is the flux density at the frequency $\nu$, $\sigma_{\rm rms}$ is the image rms noise, and N$_{\rm beam}$ is the number of beams. All flux densities in Table 4 were re-scaled to a common flux density scale using the \cite{2017ApJS..230....7P} scale and the appropriate scaling factors. The differences between the original flux density scales and the new scale are typically less than 10\% \citep{2017ApJS..230....7P}.

The total spectrum of the relic lobe is presented in Fig.~\ref{fig:spectrum}, with the new uGMRT measurements shown in magenta. The solid line represents the best-fit power-law model with a slope of $\alpha_{\rm tot} = 2.37^{+0.15}_{-0.13}$, 
consistent with the spectral index measured in G20.

\begin{table*}
\caption{Flux density of the lobe inner region and filaments}
\label{tab:flux2}
\begin{center}
\begin{tabular}{ccccccc}
\hline\noalign{\smallskip}
Region &  \multicolumn{6}{c}{Flux density (mJy)}  \\ 
       & 147 MHz & 194 MHz & 227 MHz & 333 MHz & 400 MHz & 467 MHz\\
\noalign{\smallskip}
\hline\noalign{\smallskip} 
Inner--f & $656\pm68$ & $508\pm52$ & $479\pm48$ & $213\pm11$ & $154\pm8$ &  $128\pm7$ \\
f1 & $117\pm14$ & $77\pm8$ & $66\pm7$ & $19\pm1$ & ND & ND \\

f2 & $152\pm16$ & $91\pm10$ & $67\pm7$ & $22\pm1$ & ND & ND \\

f3 & $88\pm10$ & $48\pm5$ & $33\pm3$ & $15\pm1$ & $7.9\pm0.4$ & $5.3\pm0.3$ \\
\hline{\smallskip}
\end{tabular}
\end{center}
NOTE--- Inner--f denotes the inner region with filaments f1, f3 and f3 excluded. Its total flux density was measured within the 
white polygon in Fig.~\ref{fig:ridge}. The flux density of the filaments was measured using the blue, magenta and green 
regions in Fig.~\ref{fig:ridge}. ND denotes non-detection.
\end{table*}

\begin{table*}
\caption{Spectral index of the inner region of the relic lobe and filaments.}
\label{tab:spix}
\begin{center}
\begin{tabular}{cccc}
\hline\noalign{\smallskip}
Region &  $\alpha_{\rm low}$ & $\alpha_{\rm high}$  & $\alpha_{\rm pl}$\\
       &    (147-227 MHz)  & (227-467 MHz)  & (147-467 MHz)   \\
       \noalign{\smallskip}
\hline\noalign{\smallskip} 
Inner--f  & $0.72\pm0.33$ & $1.82\pm0.16$ & $1.50\pm0.29$ \\
f1 & $1.32\pm0.37$ & \phantom{0}$3.25\pm0.31^{a}$  & \phantom{0}$2.37\pm0.13^{b}$ \\
f2 & $1.89\pm0.34$ & \phantom{0}$2.91\pm0.30^{a}$  & \phantom{0}$2.45\pm0.12^{b}$ \\
f3 & $2.26\pm0.33$ & $2.54\pm0.15$  & $2.47\pm0.10$\\
\hline{\smallskip}
\end{tabular}
\end{center}
NOTE---Inner--f denotes the inner region with filaments excluded. The observed spectral index values $\alpha_{\rm low}$ and $\alpha_{\rm high}$ were calculated from the flux densities in Table \ref{tab:flux2}. $\alpha_{\rm pl}$ is the slope from a single power law fit over the full frequency range. For Inner-f, we discuss the possibility of a curved spectrum in the text. $^{a}$: calculated between 227 and 333 MHz. $^{b}$: calculated between 147 and 333 MHz.
\end{table*}


\begin{figure*}
\gridline{\fig{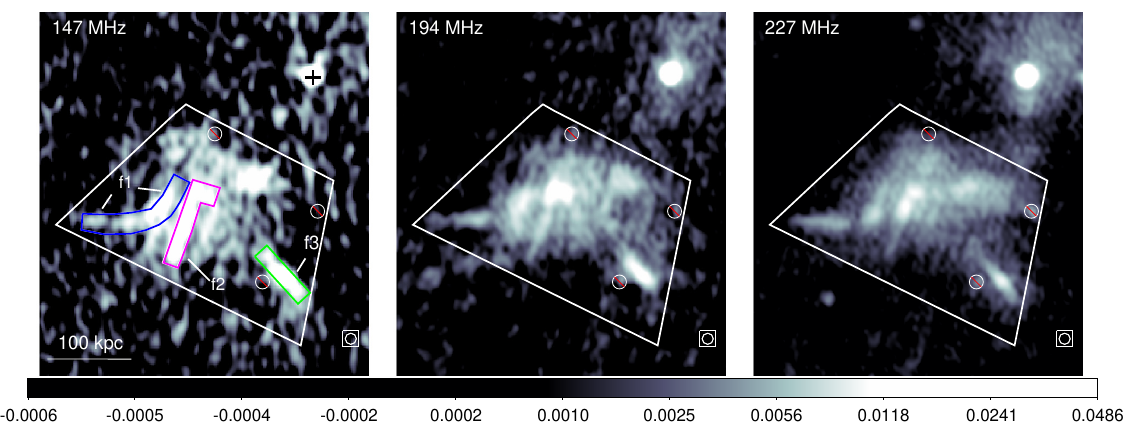}{0.9\textwidth}{}}
\gridline{\fig{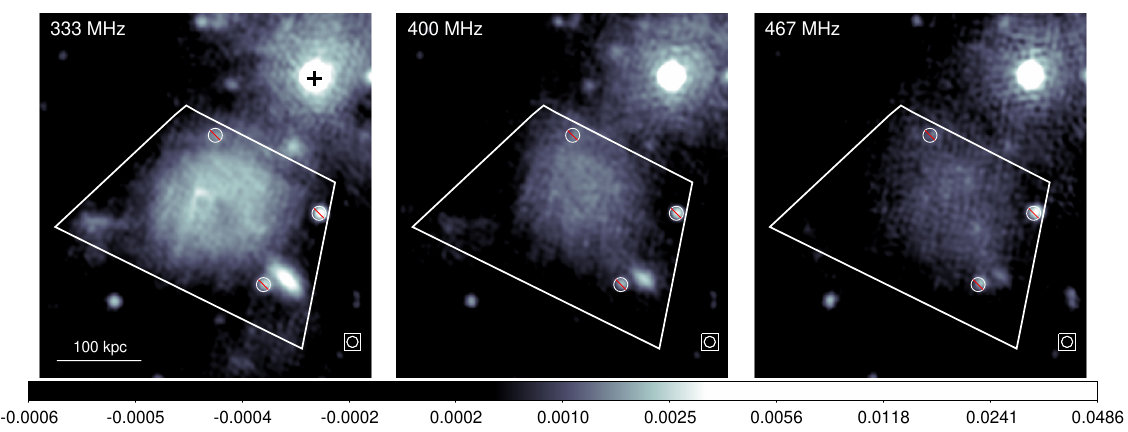}{0.9\textwidth}{}}
\caption{Images of the lobe inner region and embedded 
filaments at the 
frequencies of 147, 194, 227, 333, 400, and 467 MHz 
(images $\# 2,4,9,11,14,19$ in Tab.~3). The beam size 
is $24^{\prime\prime}$ (boxed circle in the bottom-right corner). 
The white polygon shows the ``inner'' region used to compute the spectrum 
in Fig.~\ref{fig:ridge_spectrum}(a). The blue, magenta and green regions 
were used to measure the flux density of filaments f1, f2 and f3. 
Compact sources in the lobe region have been masked out (white circles with diagonal slash). The black cross marks the BCG location. The color bar units 
are Jy beam$^{-1}$.}
\label{fig:ridge}
\end{figure*}

\begin{figure*}
\gridline{\fig{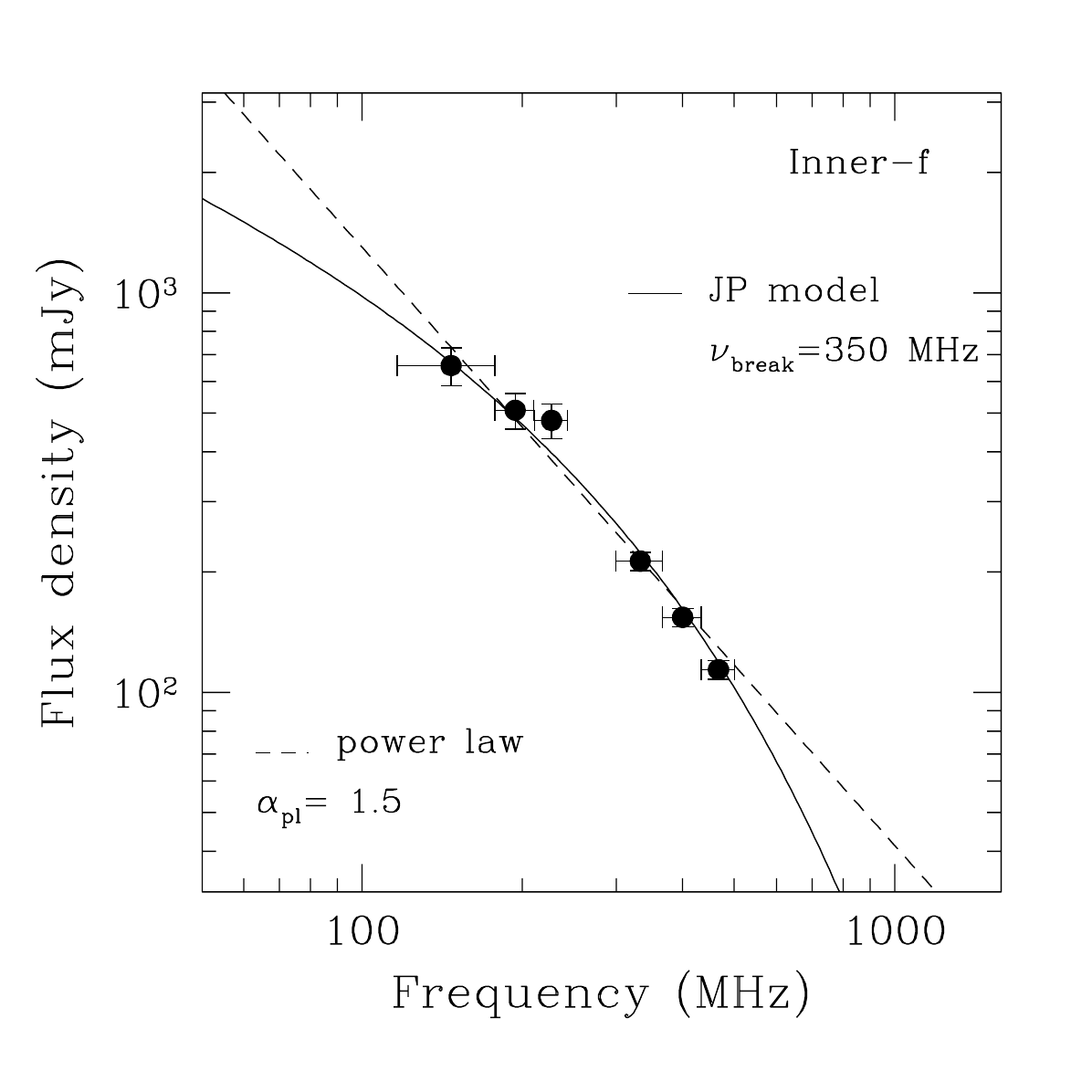}{0.5\textwidth}{(a)}
          \fig{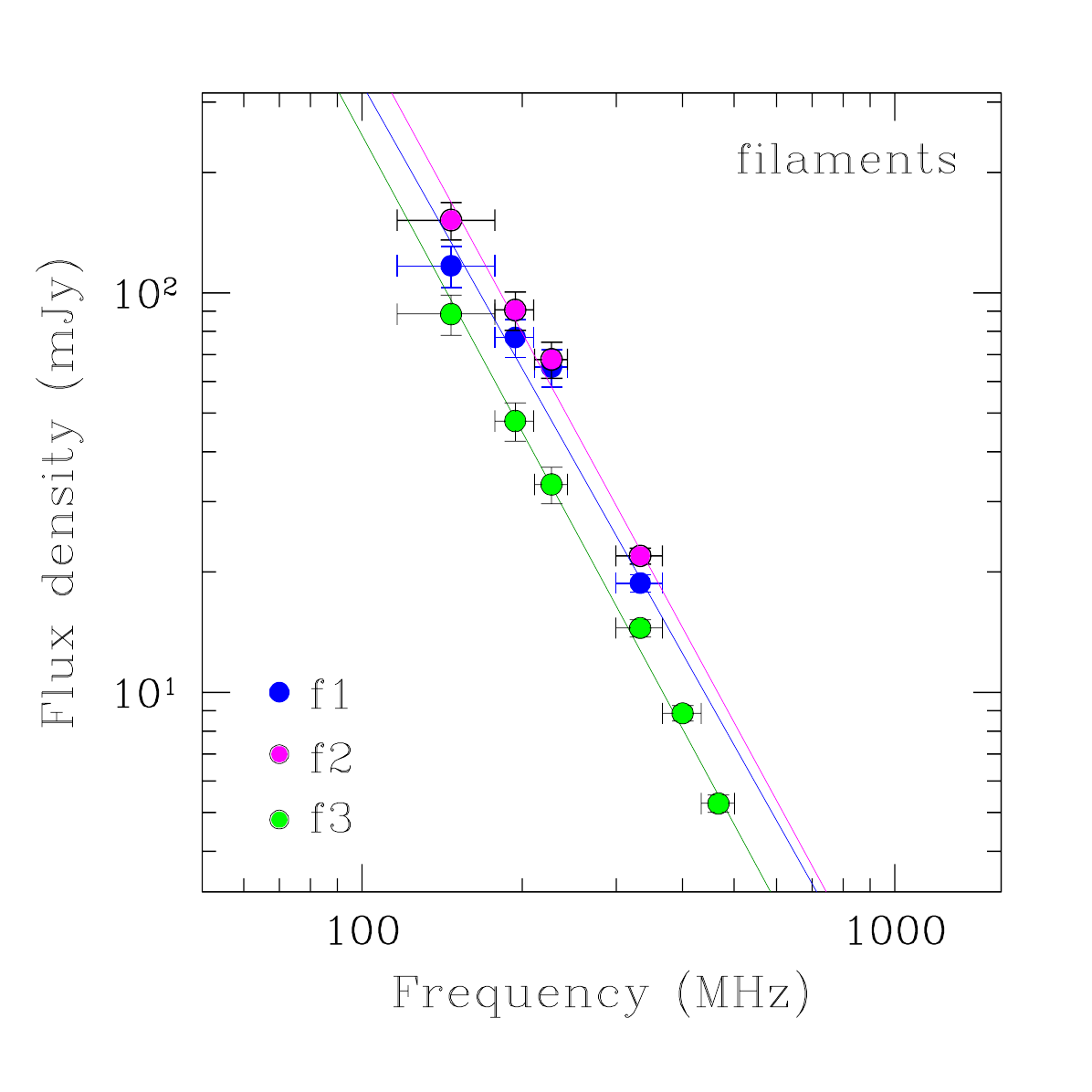}{0.5\textwidth}{(b)}}
\caption{(a) Radio spectrum of the Inner--f region of the lobe between 147 MHz and 469 MHz, computed using the white region in Fig.~\ref{fig:ridge} and excluding the filaments.
The solid curve is the best-fit JP model (see text). The best-fit break frequency from the JP is reported. The dashed line is a simple power-law fit to the data. 
(b) Radio spectrum of the filaments, computed using the flux densities in Tab.~\ref{tab:flux2} and blue, magenta and green regions in Fig.~\ref{fig:ridge}. 
The solid lines show power-law fits to the data (the corresponding best-fit slopes are summarized in Table \ref{tab:spix}).}
\label{fig:ridge_spectrum}
\end{figure*}

\subsection{Inner region and filaments}\label{sec:sp_inner}

We computed the radio spectrum of the brightest part of the lobe, located just inside the X-ray edge, utilizing a set of images at frequencies of 147, 194, 227, 333, 400, and 467 MHz, as shown in Fig.~\ref{fig:ridge}. All images were generated with a consistent $uv$ range of 0.04-14 k$\lambda$ and a resolution of $24^{\prime\prime}$ (see Table 3 for details).

The flux density was measured within the white region in Fig.~\ref{fig:ridge}. The blue, magenta, and green regions were used to measure the emission from the filaments. While the filaments are prominent in the 147 MHz image, their emission rapidly diminishes with increasing frequency. Specifically, while filament f3 is visible in all images (becoming extremely faint at 465 MHz), filaments f1 and f2 are clearly detected only in the images between 147 MHz and 333 MHz. This suggests that the filaments exhibit a significantly steeper radio spectrum compared to the underlying diffuse emission from the lobe. This is 
further supported by their lack of strong detection in new, higher-frequency MeerKAT images at 1.28 GHz \citep{botteon25}. While the MeerKAT observations are highly sensitive (4.5 $\mu$Jy beam$^{-2}$ rms at $6^{\prime\prime}$ resolution), they detect only faint emission from filament f3. Notably, the diffuse emission from the underlying innermost region of the fossil lobe is well detected by MeerKAT with a roughly circular and smooth morphology, closely resembling the structure in our highest-frequency images in Fig.~\ref{fig:ridge}.

Table \ref{tab:flux2} summarizes the flux density of the inner lobe region with filaments excluded (Inner--f) and the flux density of the filaments. For the filaments, we attempted to correct their flux densities for the local background emission from the diffuse lobe. We estimated the background level for each filament using adjacent regions of comparable size. However, the derived corrections were small, typically a few percent at most. Therefore, only uncorrected values are listed in Table \ref{tab:flux2}. 

Based on Table \ref{tab:flux2}, the filaments contribute a significant fraction of the total flux density (Inner--f + filaments) at 147 MHz ($32\%$). This contribution decreases with frequency, dropping to 26\% at 194 MHz and 19\% at 333 MHz.

Figure \ref{fig:ridge_spectrum}(a) presents the spectrum of the inner region of the lobe with filaments excluded (Inner--f). It hints at curvature, with the spectrum steepening within our frequency band --- although the band is narrow and a single power-law model (dashed line) can still qualitatively describe the data. This steepening can be quantified by comparing the spectral index in the lower ($\nu \leq 277$ MHz) and higher ($\nu \geq 277$ MHz) halves of the band. These values are given in Table~\ref{tab:spix}, and the Inner--f slope shows a statistically significant change.

To interpret this steepening in terms of the age of the electrons, we modeled the spectrum in Figure \ref{fig:ridge_spectrum}(a) using SYNAGE++ 
\citep{2011A&A...526A.148M} and adopted a Jaffe-Perola (JP) model \citep{1973A&A....26..423J}. The JP model is a widely-used aging model that assumes a single injection of relativistic electrons with a power-law energy distribution. After injection, these electrons gradually lose energy, leading to the development of a cutoff in their energy spectrum. This results in an exponential cutoff in the emitted radio spectrum at a characteristic break frequency, $\nu_{\rm br}$. The model also assumes that electron pitch angles are continuously randomized by scattering on a timescale shorter than the radiative timescale. This ensures that energy losses are statistically uniform across the electron population. Additionally, the model assumes that radiative losses dominate over other energy loss processes, and that the electron density and magnetic field are homogeneous within the emitting volume.

The solid line in Fig.~\ref{fig:ridge_spectrum}(a) represents the best-fit JP model, computed with an injection spectral index fixed at the classical value of $\alpha_{\rm inj}=0.5$ \citep[e.g.,][]{1973A&A....26..423J}\footnote{We also fit the data with $\alpha_{\rm inj}$ as a free parameter and obtained a value consistent with 0.5 within the uncertainties.}. The fit yields a break frequency of $\nu_{\rm br}=350^{+58}_{-43}$ MHz (not coincidentally, near the middle of the band where we observe the spectrum steepen). We note that the JP model with a rolling slope change is not a statistically better fit than a simple power law with $\alpha_{\rm pl}=1.5\pm0.3$ (dashed line) in this narrow  ($\sim$ 300 MHz) frequency interval, even though the change of the slope is significant, as discussed above (Table \ref{tab:spix}). Future sensitive measurements at higher frequencies ($\nu>$500 MHz) would be essential to characterize this spectrum.

\begin{figure}
\centering 
\includegraphics[width=\hsize]{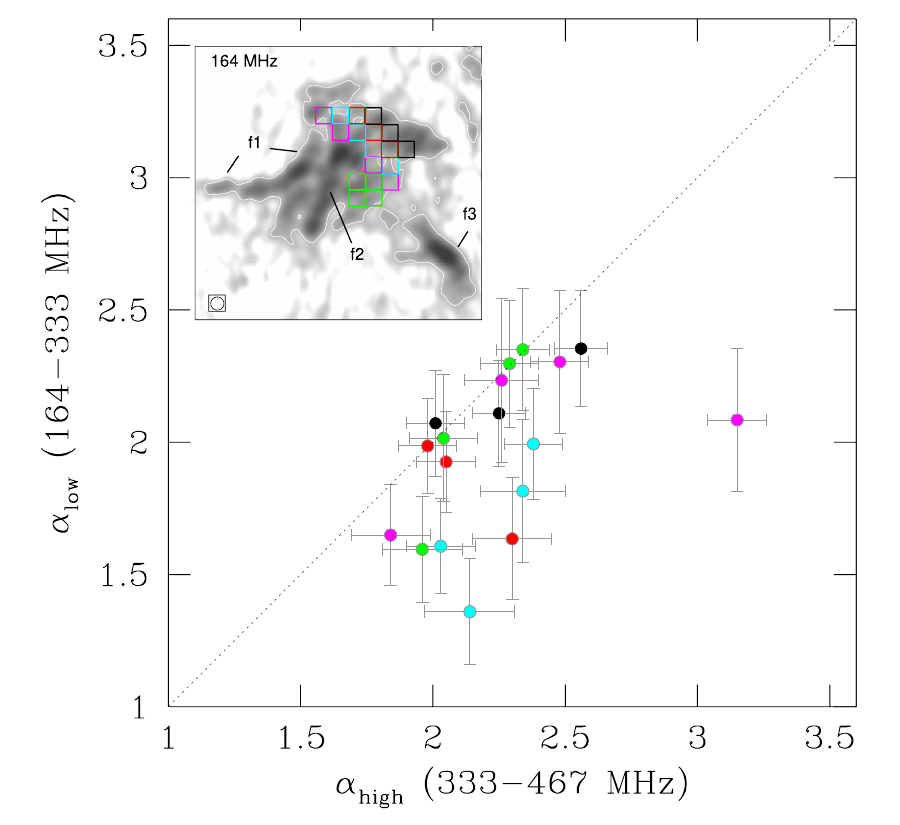}
\caption{Color-color plot for the inner lobe region. Spectral indices were derived from $24^{\prime\prime}$-resolution images at 164 MHz, 333 MHz, and 465 MHz ($\#$ 3, 10, and 19 in Tab.~3). Data points are from beam-independent regions of $30^{\prime\prime}\times30^{\prime\prime}$ and are 
color-coded as shown in the inset, with different colors denoting different distance from the cluster core.
The grid is overlaid on the 164 MHz image, where the white contour outlines the $5\sigma$ level. The dashed line represents the condition where $\alpha_{\rm  low} = \alpha_{\rm  high}$.}
\label{fig:color}

\end{figure}

To check if the spectral curvature could be dominated by some small region, we constructed a color-color diagram \citep{1997ApJ...488..146K} shown in Fig.~\ref{fig:color}. The diagram compares the spectral index between 164 MHz and 333 MHz ($\alpha_{\rm low}$) to that between 333 MHz and 467 MHz ($\alpha_{\rm high}$) for a set of beam-independent regions within the brightest part of the Inner--f region ($>5\sigma$ at 164 MHz; see inset in Fig.~\ref{fig:color}). Even though the uncertainties are large due to the proximity of the frequencies, several regions in Fig.~\ref{fig:color} fall below the power-law line (dashed line, where $\alpha_{\rm high}=\alpha_{\rm low}$), indicating that the spectral curvature is present across the region.
To investigate a potential radial trend in the curvature, we used different colors in Fig.~\ref{fig:color} to represent increasing distances from the cluster core (from black to green). However, no distinct radial pattern is observed.


The spectra of the filaments are shown in Figure~\ref{fig:ridge_spectrum}(b), and their measured spectral indices are given in Table~\ref{tab:spix}. The spectra of f1 and f2 are notably steep below 227 MHz ($\alpha_{\rm low}=1.32\pm0.37$ and $1.89\pm0.34$, respectively), and steepen even further between 227 MHz and 333 MHz ($\alpha_{\rm high}=3.25\pm0.31$ and $2.91\pm0.30$), 
suggesting spectral curvature. However, due to the limited frequency coverage (147--333 MHz), their spectrum is also well fitted by a power-law model (blue and magenta solid lines in Figure~\ref{fig:ridge_spectrum}(b)) with best-fit slopes $\alpha_{\rm f1, \, pl} = 2.37\pm0.13$ and $\alpha_{\rm f2, \, pl} = 2.45\pm0.12$ (also reported in Tab.~\ref{tab:spix}). The spectrum of f3 is consistent with a single spectral index of $\alpha_{\rm f3, \, pl} = 2.47\pm0.10$ across the entire 147-469 MHz range (green solid line).

\section{Spectral index images}\label{sec:spix}

We studied the spatial distribution of the radio spectral index within the filaments and across the lobe at both high and low angular resolutions. This was achieved by comparing pairs of images at different frequencies, using matching beams and an identical $uv$ range to ensure sensitivity to the same range of spatial scales. The spectral index was calculated for each pixel where both images exhibited a surface brightness exceeding the $3\sigma$ level. 

\subsection{Filaments}\label{sec:spix_fila}

The filaments are undetected (or become very faint) above 333 MHz (Fig.~\ref{fig:ridge} and Tab.~5). 
To study their spatial spectral index distribution, we combined a 333 MHz image ($\#10$ in Table 3) with an image at 164 MHz ($\#3$ in Table 3), obtained through joint deconvolution of sub-bands B2-01, B2-02, and B2-03 (Tab.~2). 

The resulting spectral index image and associated uncertainty map, both at a resolution of $24^{\prime\prime}$, are shown in Figs.~\ref{fig:spix1}(a) and (b). The filaments clearly stand out as regions of steepest spectral index, with $\alpha$ values as high as $\sim 3.3\pm0.1$. In contrast, the underlying diffuse emission in the lobe exhibits spectral index values ranging from $\sim 1.2\pm0.4$ to $\sim 2.5\pm0.1$.  

To better visualize the overlapping spectral components within the 164 MHz to 333 MHz frequency range in Fig.~\ref{fig:spix1}(a), we generated a series of spectral tomography images. Following \cite{1997ApJ...488..146K}, each image was calculated as

\begin{equation}
I_{\rm tom}(\alpha_{\rm t}) = I_{\rm 333 \, MHz} - \left(\frac{333}{164}\right )^{\alpha_{\rm t}} \times I_{\rm 164 \, MHz}
\end{equation}

where $I_{\rm 333 \, MHz}$ and $I_{\rm 164 \, MHz}$ are the images at 333 MHz and 164 MHz respectively and with $\alpha_{\rm t}=$ ranging from 1.5 to 3.0. In the derived tomography images, components with a spectral index steeper than $\alpha_{\rm t}$ will appear as regions of positive residual flux density, whereas features with a spectral index flatter than $\alpha_{\rm t}$ will exhibit negative residuals. Components with a spectral index matching $\alpha_{\rm t}$ will become indistinguishable from the local background. 

Tomography images for $\alpha_{\rm t}=$ 1.5, 2.0, 2.5 and 3.0 are presented in Figure \ref{fig:tomo}. The filaments appear as prominent positive features in all maps up to $\alpha_{\rm t}=2.5$. Some filament regions, particularly along f1 and f2 and near the end of f3, still exhibit positive residuals at $\alpha_{\rm t}=3.0$, suggesting a steeper spectral index, as also seen in Fig.~\ref{fig:spix1}(a). The diffuse lobe emission shows primarily positive or null residuals in the tomography images at $\alpha_{\rm t}=$ 1.5 and 2.0,  transitioning to mostly negative residuals at $\alpha_{\rm t}=2.5$. This behavior suggests an overall spectral index for the lobe emission within the range of 1.5 to 2.5.

\subsection{Fossil lobe}

\begin{figure*}
\gridline{\fig{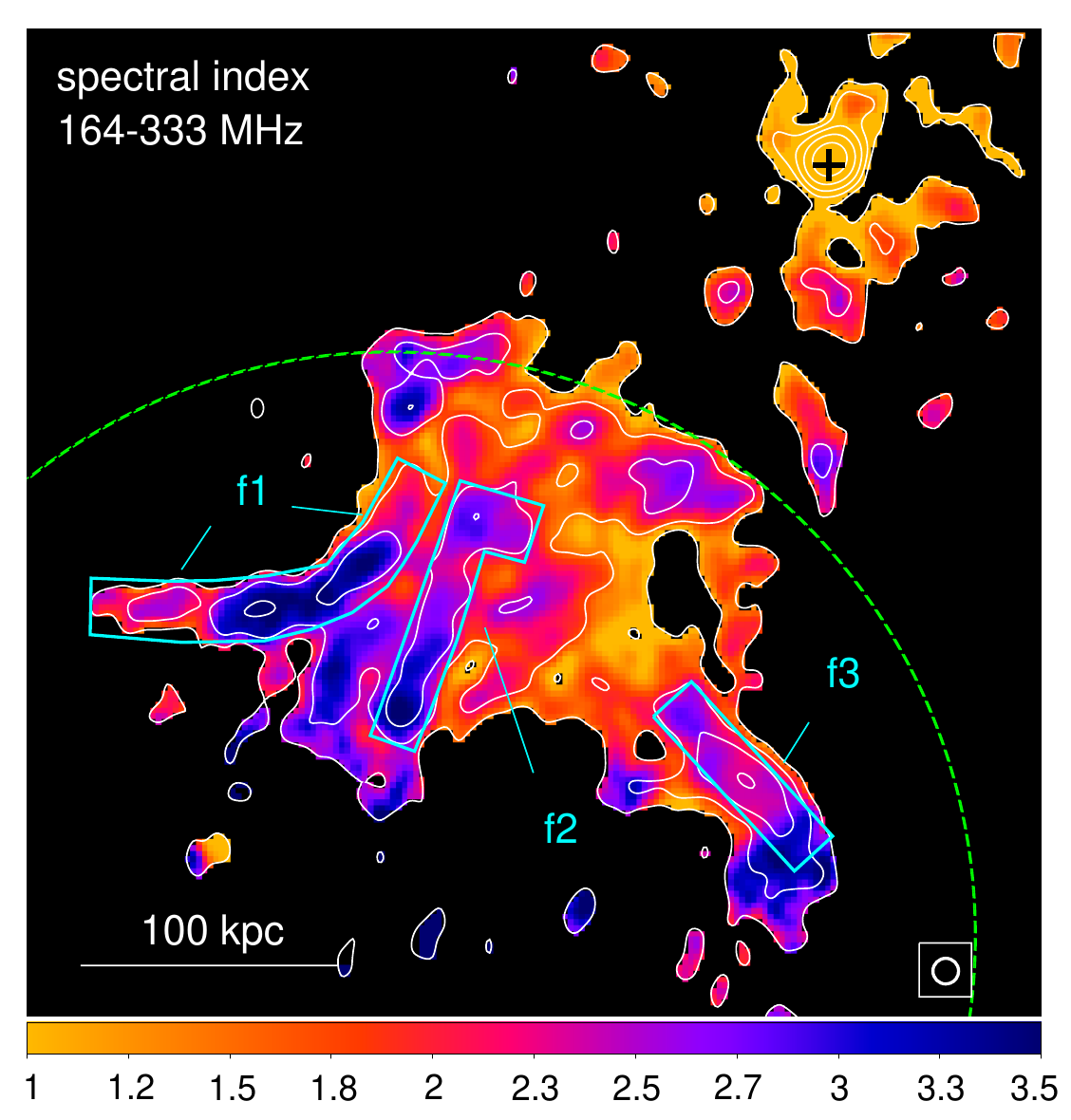}{0.45\textwidth}{(a)}
          \fig{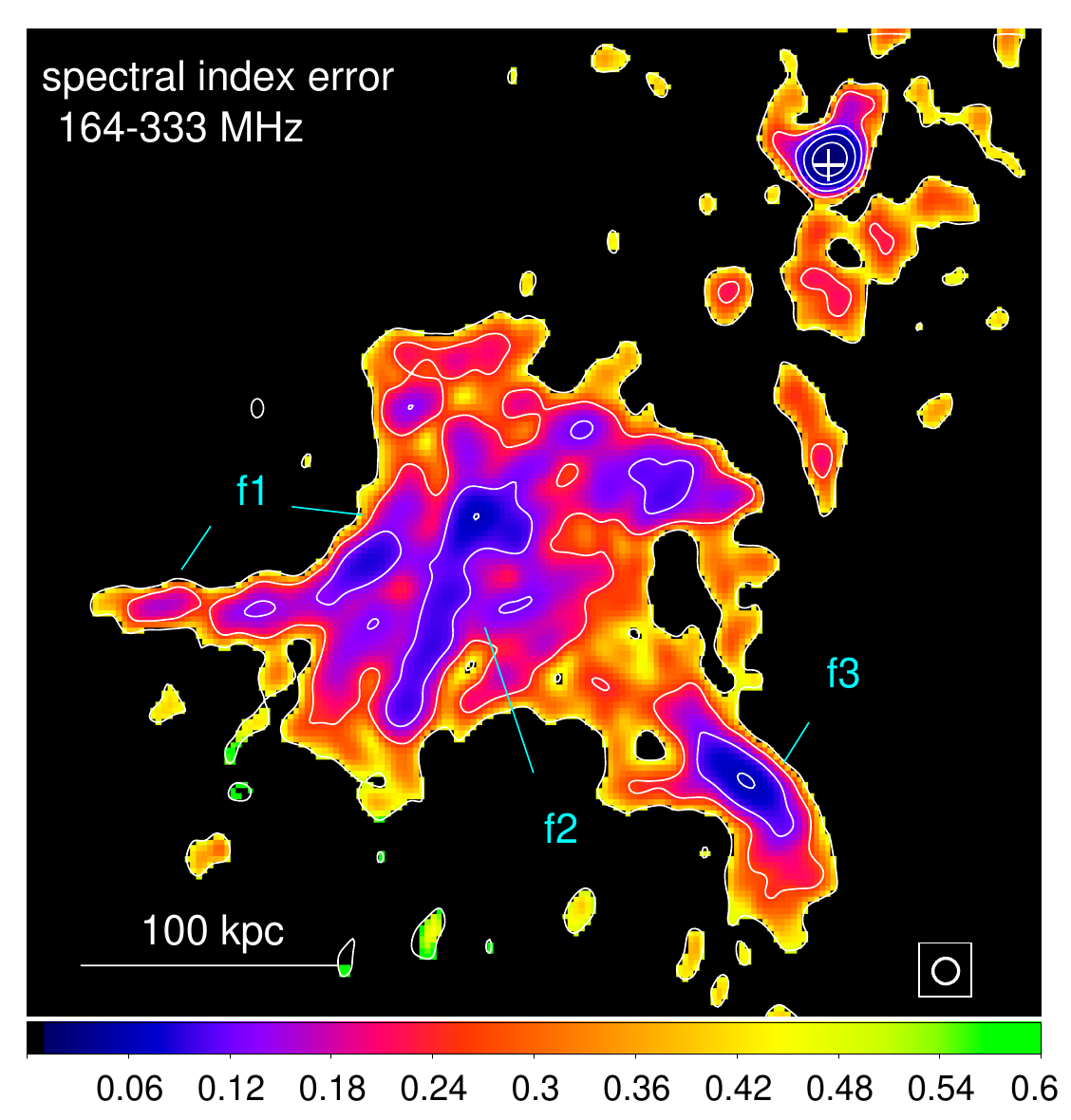}{0.45\textwidth}{(b)}}
\gridline{\fig{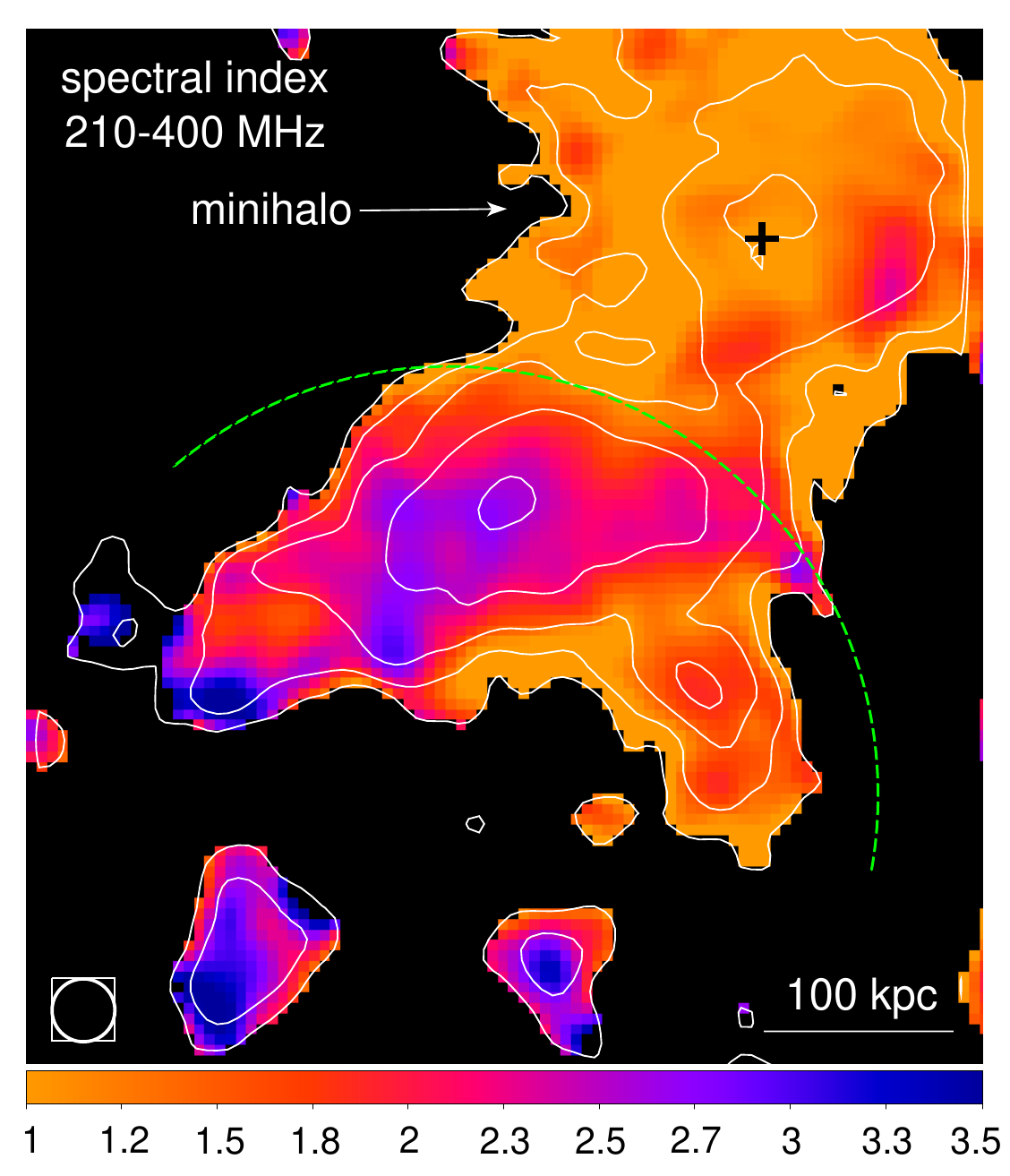}{0.45\textwidth}{(c)}
          \fig{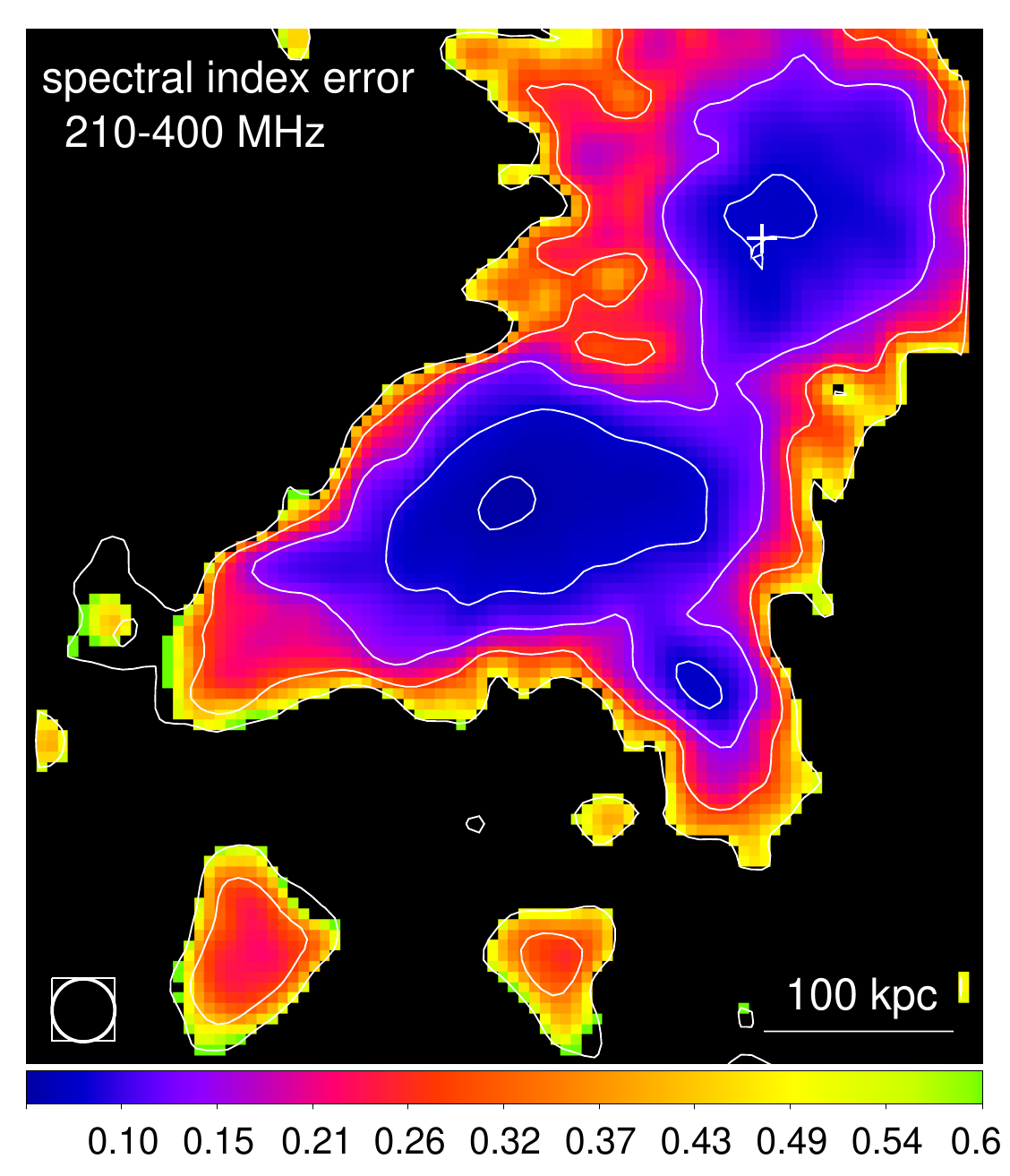}{0.45\textwidth}{(d)}}    
          
\caption{(a) Spectral index image between 164 MHz and 333 MHz 
at $24^{\prime\prime}$ resolution and (b) associated uncertainty map.
Contours at 164 MHz are shown, spaced by a factor of 2 starting from $+3\sigma=2.7$ mJy beam$^{-1}$.
(c) Spectral index image between 210 MHz and 400 MHz at $60^{\prime\prime}$ resolution and (d) associated uncertainty map. Compact sources have been 
subtracted out. Contours at 210 MHz are shown, spaced by a factor of 2 starting from $+3\sigma=3.9$ mJy beam$^{-1}$. In all panels, the beam size is 
shown as a boxed circle. The cross marks the cluster center and the green arc traces 
the X-ray edge.} 
\label{fig:spix1}
\end{figure*}

\begin{figure*}
\centering 
\includegraphics[width=17cm]{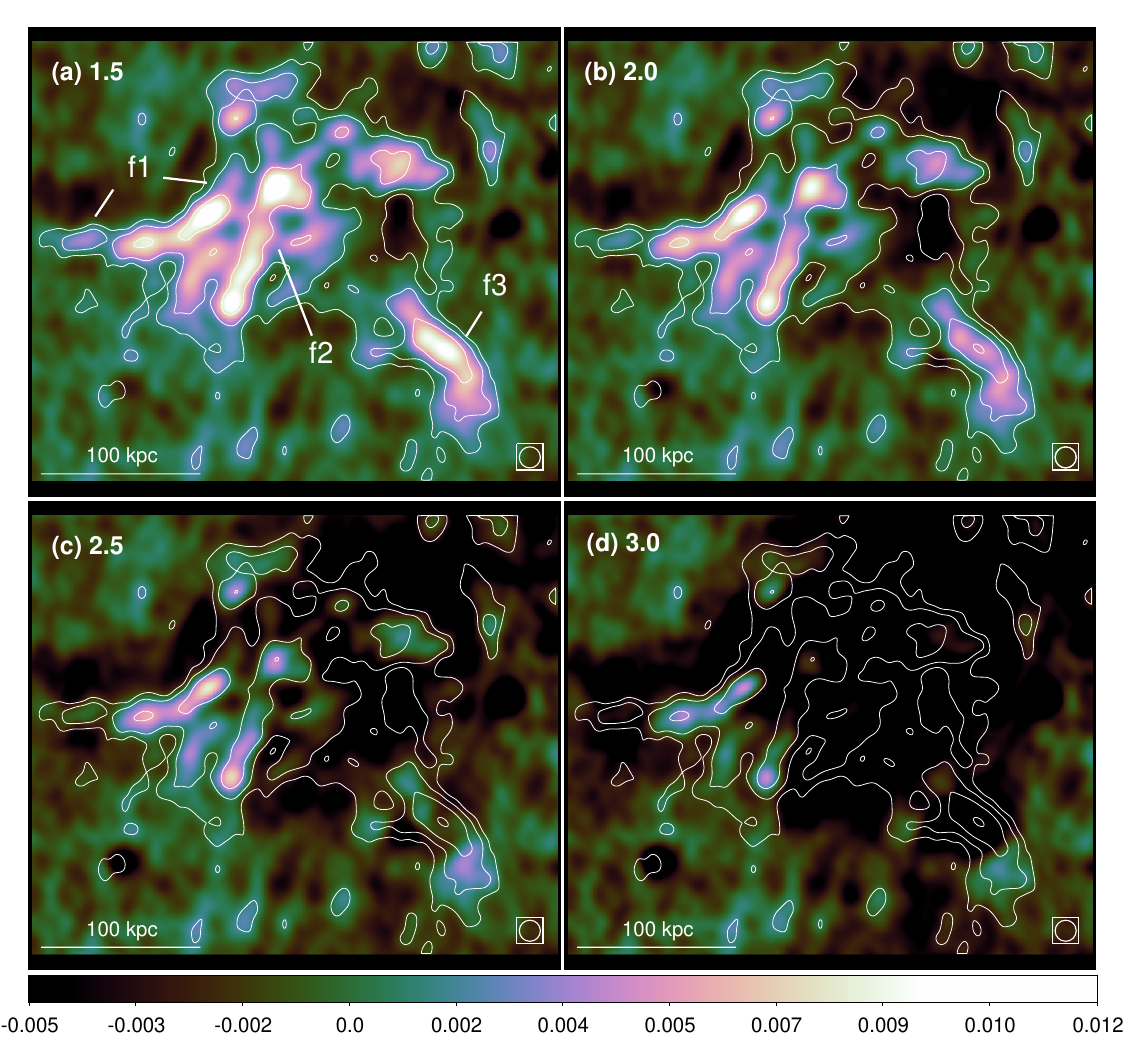}
\caption{Fossil lobe inner region and filaments: spectral tomography images between 164 MHz and 333 MHz at $24^{\prime\prime}$ resolution ($\#$ 3 and 10 in Tab.~3), derived using (a) $\alpha_{\rm t}=1.5$, (b) $\alpha_{\rm t}=2.0$, (c) $\alpha_{\rm t}=2.5$, and (d) $\alpha_{\rm t}=3.0$. Regions with $\alpha > \alpha_{\rm t}$ appear as positive residuals, while regions with a spectrum flatter than $\alpha_{\rm t}$ appear as negative residuals. Contours at 164 MHz are shown, spaced by a factor of 2 starting from $+3\sigma=2.7$ mJy beam$^{-1}$. Colors show the residual flux density and color bar units are Jy beam$^{-1}$.}
\label{fig:tomo}
\end{figure*}

Figures~\ref{fig:spix1}(c) and (d) present a spectral index image, along with associated errors, of the inner region of the fossil radio lobe derived from observations between 210 MHz and 400 MHz at a lower resolution of $60^{\prime\prime}$ and after subtraction of compact sources (images $\#$ 7 and 16 in Tab.~3). Contours at 210 MHz are overlaid at this matching resolution. The spatial distribution of spectral index within the extended emission appears complex, with no discernible spatial trends or gradients.

Overall, the spectral index fluctuates between $\alpha \sim 1.2\pm0.2$ and $\alpha \sim 3.0\pm0.3$. The steepest values are observed in the region of the filaments. Two distinct patches of steep spectrum emission, with $\alpha\sim3$, are evident at a distance of $\sim$450 kpc from the X-ray edge (green arc). These patches are located in the outermost region of the lobe, coinciding with its faintest radio emission in, e.g., Fig.~\ref{fig:400ext}.

At smaller radii, a clear separation in spectral index becomes apparent between the fossil lobe located within the X-ray edge and the diffuse radio minihalo that fills the cluster cool core. The minihalo exhibits a flatter average radio spectrum with $\alpha \sim 1.0-1.3$. This distinct spectral signature, in addition to their spatial separation (e.g., Fig.~2), reflects the different origin of these two radio emission components.

\section{Discussion}
\label{sec:disc}

\subsection{Age of the fossil lobe}\label{sec:age}

A very rough estimate of a radio source age can be inferred from the presence of a break in its synchrotron radio spectrum. Under certain assumptions, such as a constant magnetic field and negligible expansion losses, the break frequency, $\nu_{\rm br}$, 
is directly related to the time elapsed since the initial particle injection and the magnetic field strength. This relationship is given by
\begin{equation}
t_{\rm rad}= 1590\frac{B^{0.5}}{(B^2+B_{\rm IC}^2) [(1+z)\nu_{\rm br}]^{0.5}},
\label{eq:age}
\end{equation}
where $t_{\rm rad}$ is the radiative age in Myr, $\nu_{\rm br}$ is in GHz, the magnetic field intensity $B$\/ is in $\mu$G, and $B_{IC}=3.25(1+z)^2$ is the inverse Compton equivalent magnetic field in $\mu$G \citep[e.g.,][]{2011A&A...526A.148M}.

In Section \ref{sec:sp_tot}, we found that the spectrum of the entire lobe is well-described by a power law (Fig.~\ref{fig:spectrum}) with no discernible spectral breaks. This lack of a break precludes the use of spectral aging techniques to estimate the age of the radio plasma (see also discussion in G20). However, the spectrum of the innermost and brightest region of the lobe (within $\sim$200 kpc of the X-ray edge) exhibits a hint of curvature (Fig.~\ref{fig:ridge_spectrum}). In Section \ref{sec:sp_inner}, we modeled this curvature using a JP aging model. The best-fit JP model yielded a break frequency $\nu_{\rm br}=350^{+58}_{-43}$ MHz. This break frequency can then be utilized in Equation \ref{eq:age} to derive an estimate of the age for the inner region of the lobe.

To calculate the magnetic field strength, we employed the SYNAGE++ package, assuming equipartition between relativistic particles and magnetic field energy densities within the radio-emitting volume. We adopted a filling factor of 1, and equal energy contributions from relativistic ions and electrons ($k = 1$). Additionally, we imposed a low-energy cutoff ($\gamma_{\rm min} = 100$) on the energy distribution of the radio-emitting electrons \citep[e.g.,][]{1997A&A...325..898B}. This cutoff accounts for the contribution of relativistic electrons with energies as low as 50 MeV.

Using a frequency of 147 MHz and the corresponding flux density for the Inner-f region from Table \ref{tab:flux},  we calculated a radio luminosity of $P_{\rm 147 \, MHz}=1.2\pm0.1\times10^{24}$ W Hz$^{-1}$. Using this luminosity and the injection spectral index of $\alpha_{\rm inj} =0.5$, we obtained an equipartition magnetic field strength of $B_{\rm eq}=0.64\pm0.02\,\mu$G.
For this magnetic field intensity and a break frequency of $\nu_{\rm br}=350^{+58}_{-43}$ MHz, Equation \ref{eq:age} yields a radiative age $t_{\rm rad} = 174\pm15$ Myr.

We caution that this estimate is very crude. It relies on a limited frequency coverage (and the fact that the break frequency has to be in the middle of that interval simply because we observe spectral steepening) and several simplifying assumptions, including constant and uniform magnetic field across the lobe, negligible dynamical effects (from, e.g., expansion or compression), and lack of re-acceleration and mixing with ICM that may affect the energy evolution of the electrons. Deviations from these assumptions can significantly impact the estimate \citep[e.g.,][and references therein]{2002NewAR..46...95R,2017MNRAS.466.2888H}; a more comprehensive analysis that incorporates these effects is beyond the scope of the present work.

The radiative age estimated above ($\sim170$ Myr) suggests that the radio plasma within the lobe is extremely aged. This age is broadly consistent with a rough estimate of the dynamical age of the feature, $\sim 240$ Myr, derived under the assumptions that the radio lobe has started as a buoyant bubble injected near the cluster center and risen to its current location at the speed of sound (G20). 

It is interesting to compare the age of the Ophiuchus fossil lobe to similarly-derived spectral ages for other known dying/fossil radio sources. With $t_{\rm rad} \sim  170$ Myr, Ophiuchus is at the very high end of the age distribution of dying/remnant radio galaxies, for which \cite{2007A&A...470..875P} calculated a median spectral age of 63 Myr \citep[see also estimates in, e.g.,][]{2004A&A...427...79J,2007A&A...476...99G,2011A&A...526A.148M,2017A&A...600A..65S,2018A&A...618A..45B,2020A&A...634A...9M,2020MNRAS.496.3381R,2021A&A...650A.170B,2023ApJ...944..176D}.
Only a few known AGN remnant sources have
spectral ages comparable to the fossil lobe in Ophiuchus \citep[e.g.,][]{2019PASA...36...16D,2021NatAs...5.1261B,2022A&A...661A..92B,2021ApJ...909..198H,2024A&A...682A.171S}. Such long radiative lifetimes raise the question of how the rapidly fading radio emission from such systems can still be visible in the radio band. It is possible that the pressure exerted on the radio plasma by the surrounding dense ICM may limit the energy losses caused by dynamical effects (i.e., expansion of the lobes), thus prolonging the radiative lifetime of the radio plasma \citep{2011A&A...526A.148M}. In Ophiuchus, this may be the case for the innermost and brightest region of the lobe, which appears to be still relatively well confined by the ICM inside the X-ray edge, whereas the fainter outer region (stretching out to $\sim 820$ kpc from the cluster center; Fig.~\ref{fig:400ext}) is likely less confined by the gas, and thus it may be fading more rapidly and currently evolving separately from the inner region.

\subsection{Where is the counterpart lobe?}
\label{sec:counter}

As discussed in G20, in the scenario where the ultra steep-spectrum radio source is a fossil lobe from an ancient outburst of the central AGN, it is difficult to explain the lack of the counterpart lobe on the opposite side of the nucleus, seen in most radio galaxies, whether in the field or in the cluster centers. Even in our new, significantly deeper radio observations, we have not detected anything that looks like a lobe northwest of the cluster core.

As pointed out by G20,  given the very large age of the outburst (based on the observed lobe), the counterpart lobe could have propagated into a region of lower ICM density on the opposite side of the cluster, expanded and aged out of the observable radio frequency band; in this scenario, we are fortunate that the SE lobe remains detectable, likely due to its confinement by the peculiar density distribution of the ICM. The cluster also exhibits vigorous sloshing of the cool core gas \citep[][G20]{2016MNRAS.460.2752W}, possibly including very large radii \citep{markevitch25}, and 
one can imagine a scenario where one the lobes, instead of escaping the sloshing core, became entrained and completely disrupted by the tangential motions of the ICM, perhaps providing seed particles for the minihalo, as seen in hydrodynamic simulations \citep{zuhone13}. We observe bending of the lobes entrained in sloshing gas motions in many clusters, e.g., A\,2029 \citep{2013ApJ...773..114P} and Hydra A \citep{2005ApJ...628..629N}.

The new radio data offers another interesting possibility --- that the newly discovered radio bridge, seen by the uGMRT (our Fig.\ 1) and MeerKAT \citep{botteon25}, is a remnant of the counterpart lobe, after it became bent and disrupted by sloshing motions. Comparing the uGMRT images at 210 MHz and 400 MHz in Fig.\ 4, it appears that the bridge spectrum is not as steep (aged) as that of the SE relic lobe; it is possible that sloshing-induced turbulence has acted to reaccelerate the cosmic-ray electrons from the disrupted lobe \citep{zuhone13}. In this scenario, the spatial overlap of the bridge with the northern phoenix is a coincidence (indeed, the image in Fig.\ 4b suggests that the bridge does not end at the phoenix and extends past it to the northwest).

\subsection{Nature of the filaments}\label{sec:orig_fila}

Complex, filament-like features are often seen within the extended components of radio galaxies \citep[e.g.,][]{1984ApJ...285L..35P,2000ApJ...543..611O,2011MNRAS.417.2789L, 2014MNRAS.442.2867W, 2019MNRAS.488.3416H,2021ApJ...911...56G,2020A&A...634A...9M,2021NatAs...5.1261B,2024ApJ...976...64R} as well as in several diffuse radio sources in galaxy clusters \citep[e.g.,][]{2001AJ....122.1172S,2005A&A...430L...5G,2014ApJ...794...24O,2018ApJ...865...24D,2020A&A...636A..30R,2022A&A...657A..56K,2022SciA....8.7623B}. Very faint and narrow synchrotron threads have also been imaged emerging from, or in the vicinity of, the lobes and tails of a number of radio galaxies \citep[e.g.,][]{2014ApJ...794...24O,2016MNRAS.459..277S,2017SciA....3E1634D,2019A&A...627A.176C,2021ApJ...911...56G,2021A&A...649A..37B,2022MNRAS.509.1837P,2022A&A...661A..92B,2023A&A...677A...4C} with recent striking examples unveiled by deep MeerKAT observations \citep{2020A&A...636L...1R,2021ApJ...917...18C,2022A&A...657A..56K,2022ApJ...935..168R,2022ApJ...934...49G}.

The nature of these filaments is not fully understood. They could be tracing variations in the magnetic field within the radio lobes and/or a result of spatially nonuniform distribution of relativistic electrons \citep[e.g.,][]{2013MNRAS.433.3364H}. Magneto-hydrodynamic (MHD) turbulence and plasma instabilities within radio lobes may also contribute to the filamentary structure of synchrotron emission \citep[e.g.,][]{1989AJ.....98..256E,2013MNRAS.433.3364H,2013A&A...558A..19W,1995ApJ...453..332J}.
When observed in the surroundings of cluster radio galaxies, the filaments may be illuminating magnetic field structures within the cluster ambient gas \citep[e.g.,][]{2012MNRAS.422..704P,2015IAUGA..2258369B,2018SSRv..214..122D, 2020A&A...636L...1R, 2021ApJ...917...18C,2022ApJ...934...49G,2022ApJ...935..168R} or mapping ICM regions enriched with cosmic-ray electrons advected from AGN lobes and tails by gas flows driven by ``ICM weather'' \cite[e.g.,][]{2020A&A...643A.172I,2022A&A...661A..92B,2022ApJ...934...49G}; see also numerical simulations by \cite{2021ApJ...914...73Z} and \cite{2021A&A...653A..23V}.

In the Ophiuchus cluster, the radio filaments appear as narrow (5-10 kpc) and long (70-100 kpc) strands embedded within the inner region of the fossil lobe. Their radio peaks have no associated optical galaxies, which disfavors an interpretation of these features as head-tail radio galaxies superimposed on the lobe emission.
We need to explain two facts: the filaments are (a) much brighter at low frequencies that the ambient synchrotron emission from the lobe, and (b) their radio spectral index is very steep ($\alpha \sim$ 2--3), and, importantly, significantly steeper than that of the ambient emission ($\alpha \sim$ 1.5-2). The spectral slope difference contrasts with observations in other radio galaxies with filaments, where the filaments often exhibit similar or flatter spectra compared to the ambient plasma \citep[e.g.,][]{2003MNRAS.342..399G,2022A&A...658A...5T,2023A&A...677A...4C}. A striking example is the radio galaxy in Nest200047 \citep{2021NatAs...5.1261B}, whose remnant steep-spectrum radio lobes ($\alpha\sim2.5$) contain filaments with a much flatter spectral index ($\alpha\sim 0.8$). This flattening was interpreted as possible evidence for mild compression (e.g., caused by weak shocks from an AGN outburst or motions within the external medium) of the radio-emitting plasma. Compression of the relativistic electrons trapped in the magnetic field would indeed increase the synchrotron luminosity and shift the spectral cut-off to higher frequencies compared to simple radiative aging predictions, making the spectrum flatter initially.

However, an elevated magnetic field would make the radiative aging faster (because the synchrotron emissivity is higher), which moves the exponential cutoff to lower frequencies. The steeper slope of the Ophiuchus filaments, relative to the ambient spectrum, seems to exclude recent compression or re--acceleration of the ambient electron pool as the origin of the filaments, but a relatively faster aging at the locations of stronger $B$\/ fields offers a natural physical mechanism for the differences in spectral shapes. Filaments of ordered, elevated $B$\/ fields can form as a result of MHD turbulence, which creates regions where the magnetic field is folded, stretched and amplified \citep[e.g.,][]{2004ApJ...601..778T, 2004ApJ...612..276S, 2013MNRAS.433.3364H}. Such turbulence in the lobe can be driven by active jets (probably not a viable mechanism in the very old Ophiuchus bubble, where jet activity has presumably terminated long ago) or by circulation of the gas inside a buoyantly rising bubble. 

There are timescales that should match for this explanation to work --- namely, the radiative aging timescale and the lifetime of such filamentary field structures in the turbulent medium. A notable case with filaments in a lobe of the radio galaxy, qualitatively similar to those in Ophiuchus, is Centaurus A \citep[e.g.,][]{2011ApJ...740...17F}, with the spectral index of the filaments  ($\alpha \sim 0.8$) somewhat steeper than that of the ambient emission ($\alpha \sim 0.5-0.7$) \citep{2014MNRAS.442.2867W}. These authors conclude that it is difficult to match those timescales for Centaurus A without invoking very high magnetic field strengths. Any turbulence should also not result in the disruption of the bubble.

The Ophiuchus giant lobe is a combination of puzzles, some of which were already identified in G20 and include its origin (how did such an extraordinarily powerful outburst not destroy the cluster cool core?) and its survival (how did the bubble rise high above the cluster cool core while remaining a coherent structure?), to which we now add the discovery of filaments in the lobe with extremely steep synchrotron spectra. These puzzles are probably related, and it would thus be valuable to create a numerical MHD model of an evolving giant buoyant bubble that reproduces all these observations, which may help constrain the physical processes involved in such an extreme episode of the AGN feedback.

\section{Summary}

We presented a deep radio follow-up of the giant fossil radio lobe in the Ophiuchus cluster using high-sensitivity observations with the uGMRT in Bands 2 (125--250 MHz) and 3 (300--500 MHz). These new observations enabled us to trace the faint diffuse emission from the relic lobe to a remarkable distance of approximately 820 kpc from the cluster center. 

The new images unveiled intricate spatial structures within the fossil lobe, including the discovery of narrow (5-10 kpc), elongated (70-100 kpc) radio filaments embedded within the lower surface brightness emission of the lobe. These filaments exhibit a very steep spectral index ($\alpha \sim 3$), significantly steeper than the underlying smoother radio emission from the lobe ($\alpha \sim1.5-2$). Their origin remains uncertain. They may represent regions where the magnetic field has been stretched and amplified by MHD turbulence within the rising bubble, which would result in accelerated aging of the relativistic electrons and steepening of their synchrotron spectrum.

The spectrum of the brightest region of the fossil lobe, closest to the cluster core, displays a spectral break, from which we inferred a radiative age of approximately 174 Myr, 
suggesting the date of this AGN outburst. However, this age estimate relies 
on a limited frequency coverage (i.e., from 147 to 467 MHz). Future flux density measurements at higher frequencies will be essential to more accurately constrain the spectral curvature and to perform resolved spectral fitting across the fossil lobe. 

A clear fossil lobe counterpart from this ancient AGN outburst remains undetected. However, the new uGMRT images, supported by MeerKAT observations at 1.28 GHz \citep{botteon25}, reveal a faint radio radio bridge northeast of the cluster core. It is possible that this emission represents the remnant of the counterpart lobe that was bent and highly disrupted by gas sloshing motions.
\\
\\
Ongoing high-sensitivity surveys at frequencies around 1 GHz and below, conducted by instruments such as the LOw Frequency ARray (LOFAR), MWA, MeerKAT, and Australia SKA Pathfinder (ASKAP), along with future sensitive radio telescopes like the Square Kilometre Array (SKA), are well-positioned to uncover more of these radio fossils. The associated cavities in the ICM will be challenging to detect. These cavities are typically large and exhibit very low X-ray surface brightness contrast, especially when located outside the bright central core of the cluster. Future X-ray missions, such as NASA's proposed Advanced X-ray Imaging Satellite (AXIS) Probe, may bring the necessary combination of high angular resolution and collecting area to detect these elusive low-contrast X-ray cavities beyond the cluster cores. Determining how frequent (or rare) such extreme events have been over the cluster history would inform us on the importance of AGN feedback for cluster evolution.

{\it Acknowledgements.}
We thank the referee for their critical and helpful comments.
Basic research in radio astronomy at the Naval Research Laboratory is supported by 6.1 Base funding. We thank the staff of the GMRT who made these observations possible.
GMRT is run by the National Centre for Astrophysics of the Tata Institute of Fundamental Research.
This research made use of hips2fits,\footnote{https://alasky.cds.unistra.fr/hips-image-services/hips2fits} a service provided by CDS.

\appendix

\section{Extended radio galaxies in the Ophiuchus field}\label{sec:app_images}

\begin{figure*}
\centering \epsscale{1.1}
\includegraphics[width=16cm]{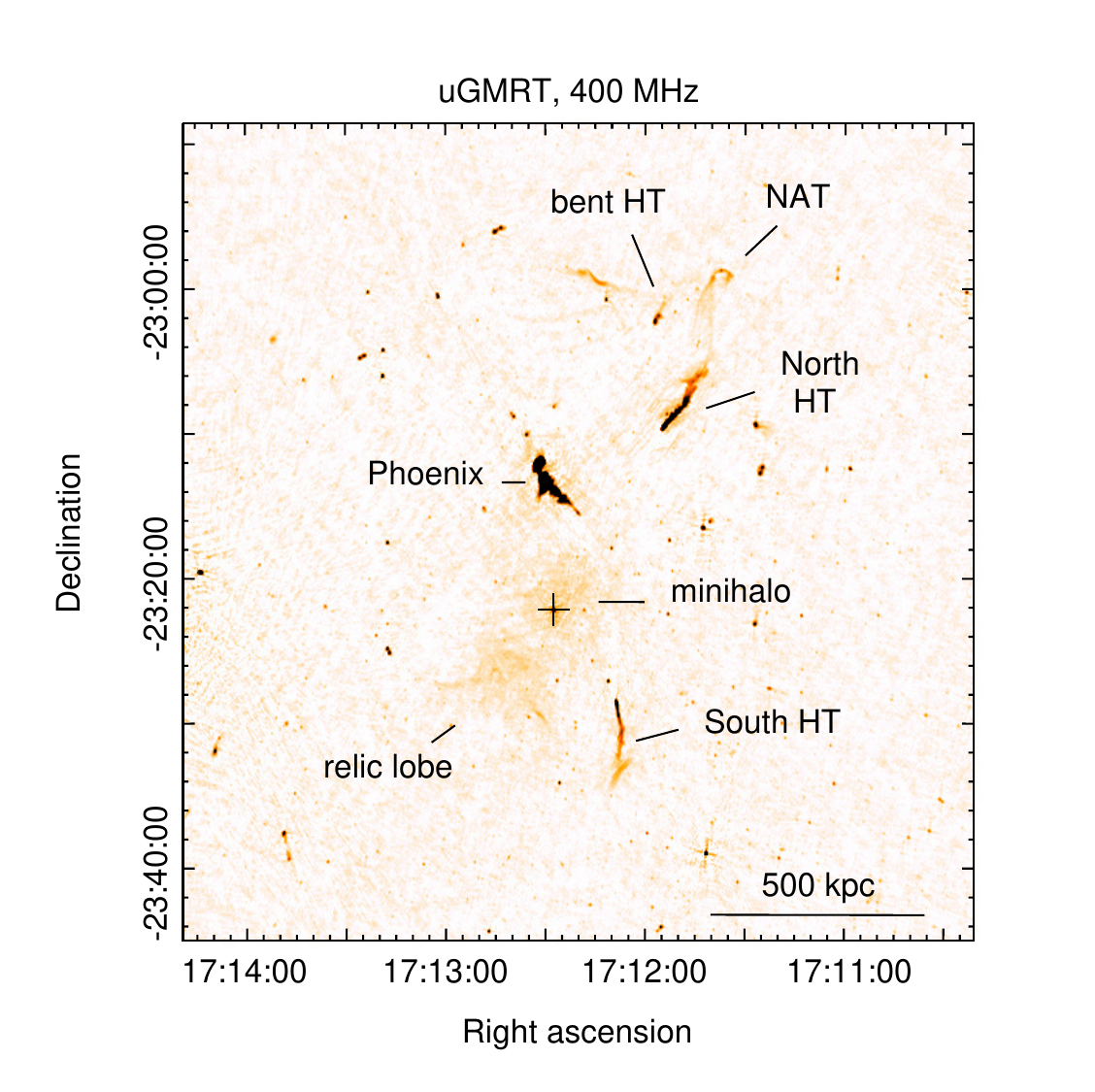}
\caption{uGMRT 400 MHz image of the Ophiuchus cluster at $9^{\prime\prime}\times6^{\prime\prime}$ resolution (image $\#12$ in Tab.~3). The $\sim 1^{\circ}\times1^{\circ}$ field ($\sim$ 2 Mpc $\times$ 2 Mpc) is fully within the GMRT Band 3 primary beam (full width at half-maximum=$75^{\prime}$) and has an rms noise of $24$ $\mu$Jy beam$^{-1}$. The image is not corrected for the primary beam effects. The position of the BCG at the cluster center is marked with a black cross. Sources of interest are labelled.} 
\label{fig:field}
\end{figure*}

Figure \ref{fig:field} presents a 400 MHz uGMRT radio image of the Ophiuchus field at a native resolution of $9^{\prime\prime}\times6^{\prime\prime}$ resolution, obtained using a Briggs robust of 0 (image $\#12$ in Tab.~3.). The image encompasses a region of $\sim 1^{\circ}\times1^{\circ}$, corresponding to $\sim$ 2 Mpc $\times$ 2 Mpc. A number of extended radio galaxies with highly distorted morphologies are situated within the field. Their high-resolution images at 400 MHz, obtained using uniform weights ($\#11$ in Tab.~3), are presented in Fig.~\ref{fig:tails}. A brief description of each is given below. We refer to \cite{botteon25} for recent MeerKAT 1.28 GHz images of these sources, which exhibit similarly intriguing features to those described here.

\subsection{Notes on individual radio sources} 

\begin{figure*}
\gridline{\fig{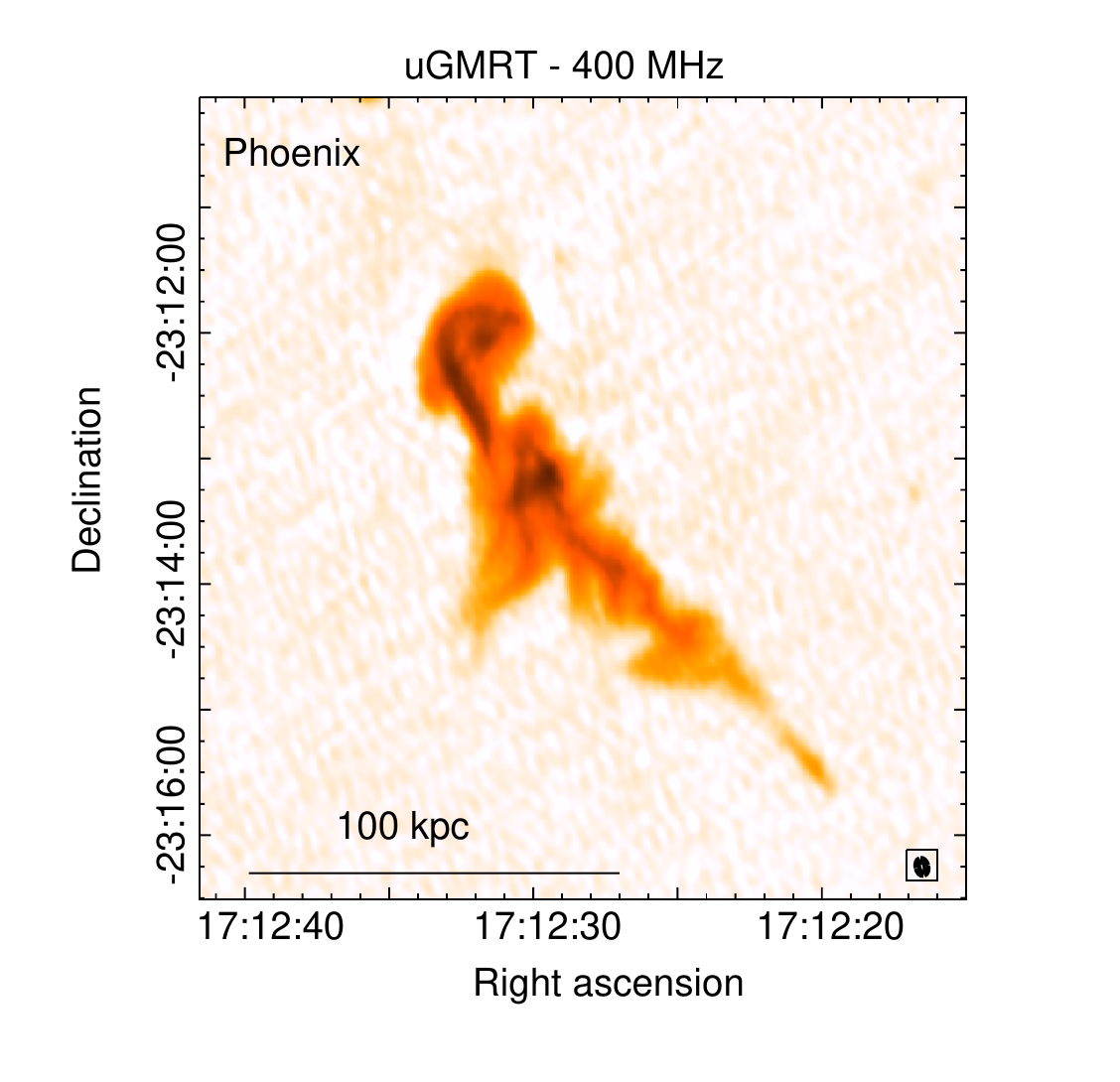}{0.5\textwidth}{(a)}
          \fig{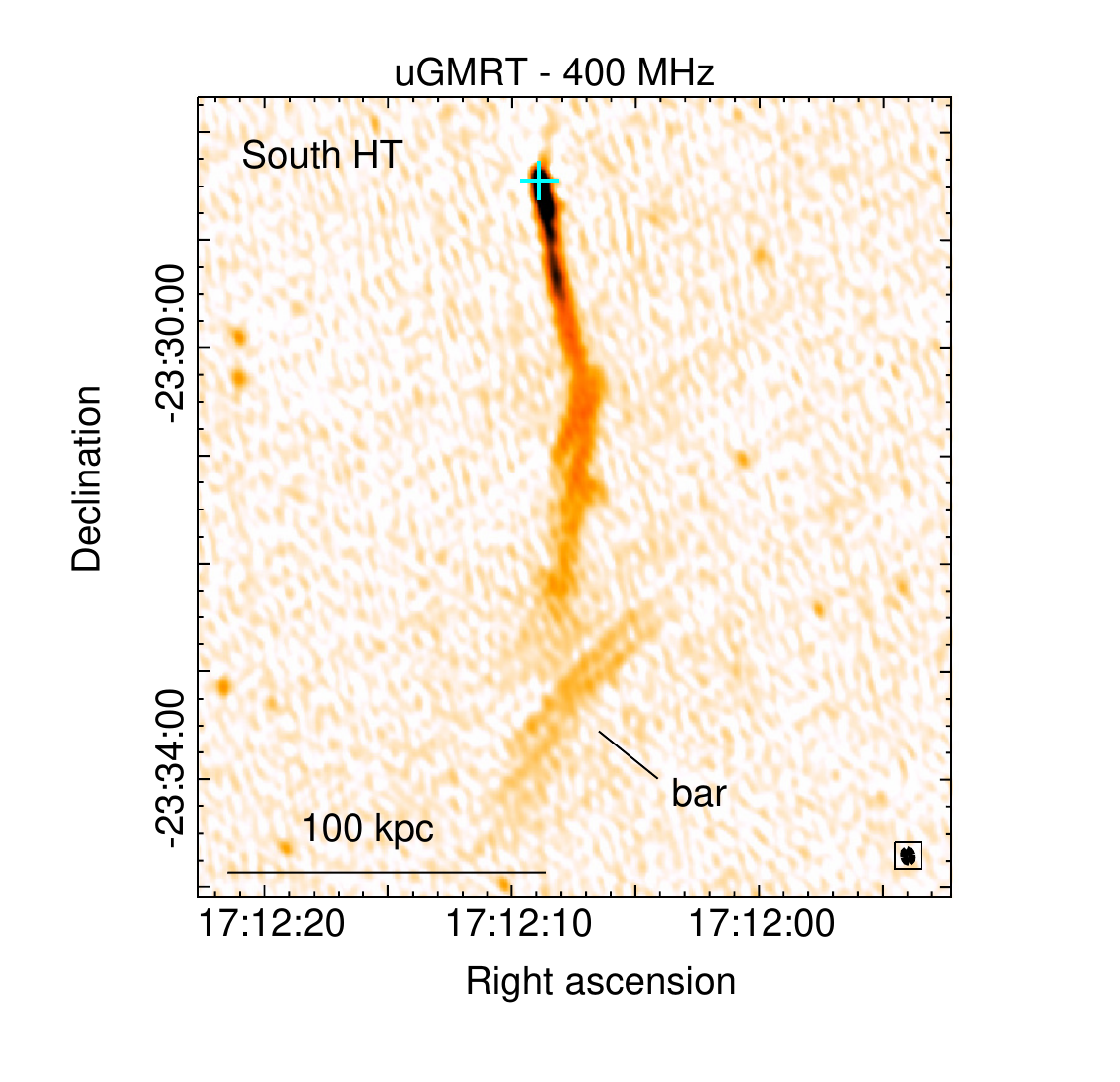}{0.5\textwidth}{(b)}}
\gridline{\fig{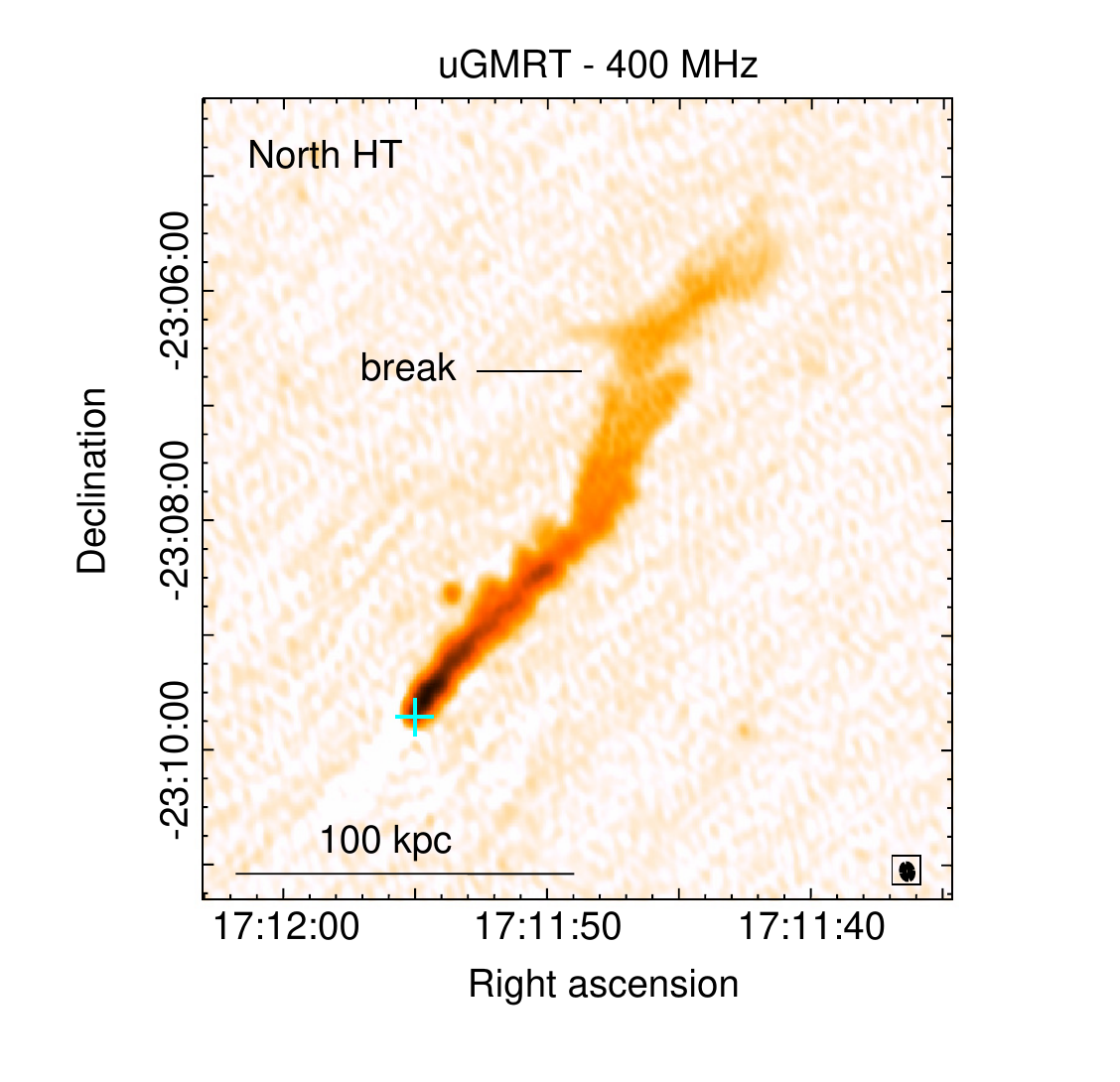}{0.5\textwidth}{(c)}
          \fig{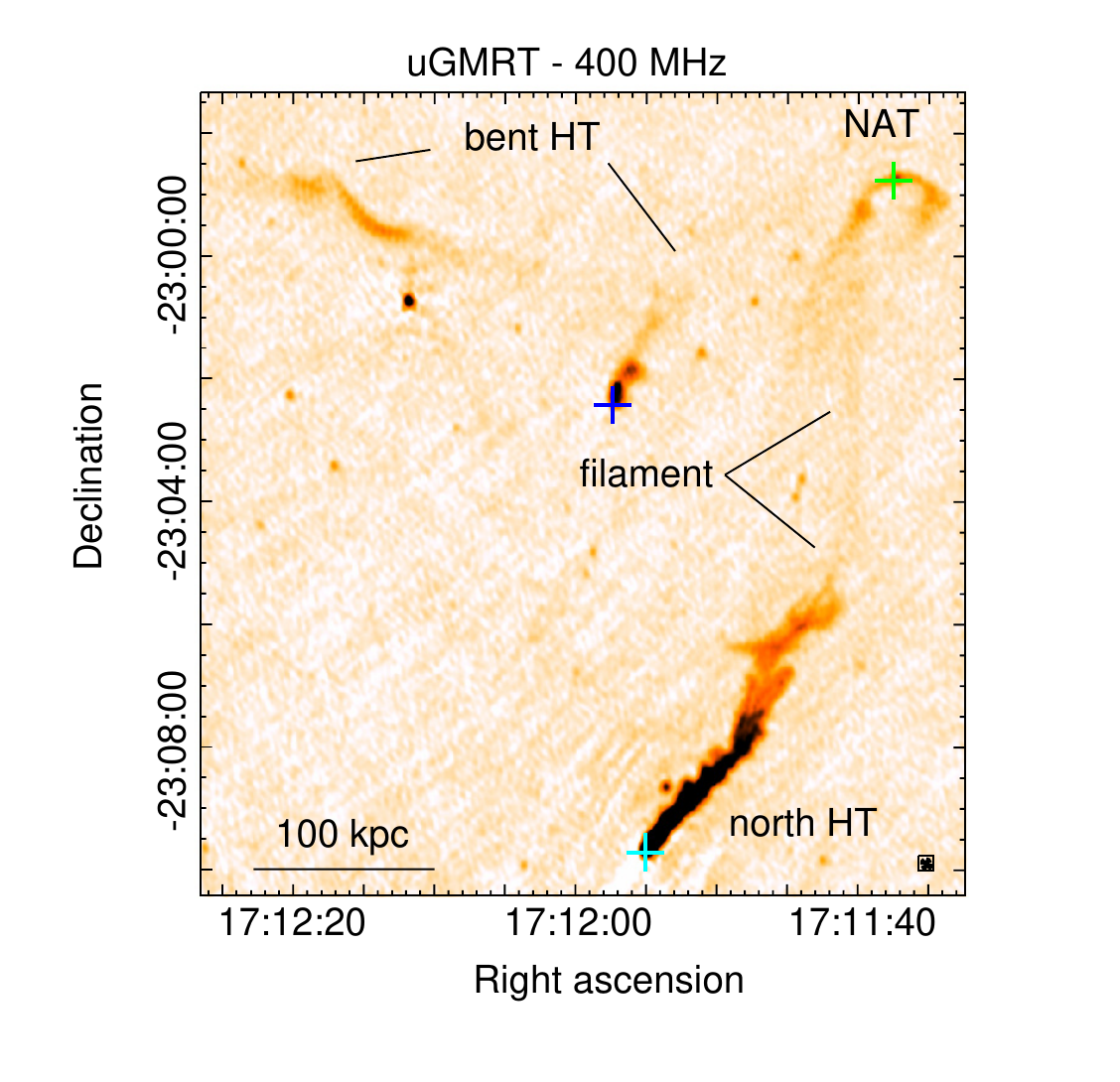}{0.5\textwidth}{(d)}
          }
\caption{uGMRT 400 MHz images of extended radio sources in the Ophiuchus field at $7^{\prime\prime}\times4^{\prime\prime}$ resolution (image $\#11$ in Tab.~3) and  $30$ $\mu$Jy beam$^{-1}$ rms noise. Cyan crosses mark the position of the optical hosts, when present.
} 
\label{fig:tails}
\end{figure*}

\smallskip\noindent{\bf Phoenix.} The radio morphology of this source in Fig.~\ref{fig:tails}(a) is highly unusual. As also observed by \cite{2010A&A...514A..76M}, \cite{2016MNRAS.460.2752W} and \cite{botteon25}, it exhibits a complex network of filaments roughly perpendicular to its major axis, along with a vortex-like structure in the northwest. It lacks a discernible radio core and has no clear counterpart in optical/near-infrared or X-ray observations \citep{2016MNRAS.460.2752W}. Its radio spectral index is very steep, $\alpha \sim 2$ \citep{2016MNRAS.460.2752W}. As proposed by \cite{2016MNRAS.460.2752W}, a plausible explanation for these characteristics involves fossil AGN plasma re-energized by adiabatic compression due to shocks or gas motions within the ICM \citep{2001A&A...366...26E}. As 
described in Sect.~\ref{sec:region}, these peculiar radio features might also be explained by a radio galaxy jet that has experienced kink instability \citep{botteon25}.

\smallskip\noindent{\bf South HT.} 
This head-tail radio source is associated with a cluster member galaxy VVVX J171209.06-232826.4 at $z=0.0249$, as reported in the NASA/IPAC Extragalactic Database (NED). Its location is indicated by the cyan cross in Fig.~\ref{fig:tails}(b). The tail extends for approximately 
220 kpc. The tail appears relatively straight and well-collimated for the first $\sim$75 kpc. Beyond this point, it bends, disrupts, and loses collimation. At a distance of $\sim$150 kpc from the head, a prominent 100 kpc bar emerges, misaligned by approximately $45^{\circ}$ with respect to the tail axis. This bar appears to consist of two distinct and narrow parallel filaments, as also seen in the MeerKAT images by \cite{botteon25}. 

\smallskip\noindent{\bf North HT.} This head-tail radio source is associated with a cluster member galaxy WISEA J171155.39-230942.6 at $z=0.0268$ (NED). Its location is indicated by the cyan cross in Fig.~\ref{fig:tails}(c). The radio tail extends northwestward for a projected length of approximately 180 kpc. Between 30 and 70 kpc from the head, the tail exhibits transverse extensions resembling the "ribs" recently observed in a head-tail radio galaxy in Abell 3266 \citep{2022A&A...657A..56K} and in ESO 137–G007 in Abell 3627 \citep{kori24}. These features may arise from interactions with an inhomogeneous external medium, instabilities in the flow \citep{upreti24} or intermittently restarting jets \citep{2021Galax...9...81R}. The tail continues for about 110 kpc beyond this region, sharply bending, widening, getting dimmer and undergoing significant disruption with the emergence of several filamentary structures. 

\smallskip\noindent{\bf NAT.} This narrow-angle tail (NAT) radio source is associated with a cluster member galaxy WISEA J171137.52-225845.9 at $z=0.0287$ (NED; green cross in Fig.~\ref{fig:tails}(d)). The jets initially exhibit a relatively symmetric bending over the first $\sim 30$ kpc. Thereafter, both tails dim and bend southeast, running parallel for approximately 30 kpc. Beyond this point, they further widen and fade significantly. A striking feature is a faint, collimated filament of emission extending 150 kpc north-south, apparently connecting (in projection) the NAT's tails to the disrupted termination of the North HT \citep[see also][]{botteon25}. Recent high-quality MeerKAT observations have imaged low surface brightness radio filaments in various cluster environments, including synchrotron threads between the lobes in ESO 137–006 \citep{2020A&A...636L...1R}, emerging from jets of IC 4296 \citep{2021ApJ...917...18C}, branching out of a head-tail in A\,3562 \citep{2022ApJ...934...49G}, and departing from a lobe of 3C\,40B and between 3C\,40B and nearby radio galaxy 3C\,40A \citep{2022A&A...657A..56K,2022ApJ...935..168R}. The origin of these enigmatic synchrotron threads remains uncertain. They may be tracing magnetic field structures within the surrounding ICM.

\smallskip\noindent{\bf Bent HT.} This source, potentially associated with the galaxy WISEA J171157.23-230211.7 at $z=0.0305$ (NED; blue cross in Fig.~\ref{fig:tails}(d)), exhibits a complex morphology. It displays an inner bright head-tail (or double-lobe) structure within the initial 25 kpc, followed by a significantly fainter tail extending approximately 40 kpc northwest. Beyond this point, the tail abruptly bends 90 degrees eastward, becoming extremely faint before re-brightening into a twisted, wiggling structure at a distance of 150 kpc from the bend. The overall extent of the source, measured from the optical host to the end of the twisted structure, is approximately 300 kpc.

{}


\begin{thebibliography}{}


\bibitem[Biava et al.(2021)]{2021A&A...650A.170B} Biava, N., Brienza, M., Bonafede, A., et al.\ 2021, \aap, 650, A170. doi:10.1051/0004-6361/202040063

\bibitem[Birkinshaw \& Worrall(2015)]{2015IAUGA..2258369B} Birkinshaw, M. \& Worrall, D.\ 2015, IAU General Assembly


\bibitem[Botteon et al.(2021)]{2021A&A...649A..37B} Botteon, A., Giacintucci, S., Gastaldello, F., et al.\ 2021, \aap, 649, A37. doi:10.1051/0004-6361/202040083


\bibitem[Botteon et al.(2022)]{2022SciA....8.7623B} Botteon, A., van Weeren, R.~J., Brunetti, G., et al.\ 2022, Science Advances, 8, eabq7623. doi:10.1126/sciadv.abq7623


\bibitem[Botteon et al.(2025)]{botteon25} Botteon, A., Balboni, M., Bartalucci, I., et al.\ 2025, \aap, MeerKAT L-band observations of the Ophiuchus galaxy cluster: Detection of synchrotron threads and jellyfish galaxies, 698, A55. doi:10.1051/0004-6361/202554695

\bibitem[Brienza et al.(2018)]{2018A&A...618A..45B} Brienza, M., Morganti, R., Murgia, M., et al.\ 2018, \aap, 618, A45

\bibitem[Brienza et al.(2021)]{2021NatAs...5.1261B} Brienza, M., Shimwell, T.~W., de Gasperin, F., et al.\ 2021, Nature Astronomy, 5, 1261

\bibitem[Brienza et al.(2022)]{2022A&A...661A..92B} Brienza, M., Lovisari, L., Rajpurohit, K., et al.\ 2022, \aap, 661, A92

\bibitem[Briggs(1995)]{1995PhDT.......238B} Briggs, D.~S.\ 1995, Ph.D. Thesis, New Mexico Institute of Mining and Technology

\bibitem[Brunetti et al.(1997)]{1997A&A...325..898B} Brunetti, G., Setti, G., \& Comastri, A.\ 1997, \aap, 325, 898 doi:10.48550/arXiv.astro-ph/9704162

\bibitem[Candini et al.(2023)]{2023A&A...677A...4C} Candini, S., Brienza, M., Bonafede, A., et al.\ 2023, \aap, 677, A4. doi:10.1051/0004-6361/202347036

\bibitem[Clarke et al.(2019)]{2019A&A...627A.176C} Clarke, A.~O., Scaife, A.~M.~M., Shimwell, T., et al.\ 2019, \aap, 627, A176. doi:10.1051/0004-6361/201935584

\bibitem[Cohen \& Clarke(2011)]{2011AJ....141..149C} Cohen, A.~S. \& Clarke, T.~E.\ 2011, \aj, 141, 149. doi:10.1088/0004-6256/141/5/149

\bibitem[Condon et al.(1998)]{1998AJ....115.1693C} Condon, J.~J., Cotton, W.~D., Greisen, E.~W., et al.\ 1998, \aj, 115, 1693. doi:10.1086/300337

\bibitem[Condon et al.(2021)]{2021ApJ...917...18C} Condon, J.~J., Cotton, W.~D., White, S.~V., et al.\ 2021, \apj, 917, 18. doi:10.3847/1538-4357/ac0880



\bibitem[de Gasperin et al.(2017)]{2017SciA....3E1634D} de Gasperin, F., Intema, H.~T., Shimwell, T.~W., et al.\ 2017, Science Advances, 3, e1701634. doi:10.1126/sciadv.1701634


\bibitem[Di Gennaro et al.(2018)]{2018ApJ...865...24D} Di Gennaro, G., van Weeren, R.~J., Hoeft, M., et al.\ 2018, \apj, 865, 24. doi:10.3847/1538-4357/aad738



\bibitem[Donnert et al.(2018)]{2018SSRv..214..122D} Donnert, J., Vazza, F., Br{\"u}ggen, M., et al.\ 2018, \ssr, 214, 122. doi:10.1007/s11214-018-0556-8



\bibitem[Duchesne \& Johnston-Hollitt(2019)]{2019PASA...36...16D} Duchesne, S.~W. \& Johnston-Hollitt, M.\ 2019, \pasa, 36, e016. doi:10.1017/pasa.2018.26

\bibitem[Dutta et al.(2023)]{2023ApJ...944..176D} Dutta, S., Singh, V., Chandra, C.~H.~I., et al.\ 2023, \apj, 944, 176. doi:10.3847/1538-4357/acaf01

\bibitem[Eilek(1989)]{1989AJ.....98..256E} Eilek, J.~A.\ 1989, \aj, 98, 256

\bibitem[En{\ss}lin \& Gopal-Krishna(2001)]{2001A&A...366...26E} En{\ss}lin, T.~A. \& Gopal-Krishna\ 2001, \aap, 366, 26

 \bibitem[Fabian(2012)]{2012ARA&A..50..455F} Fabian, A.~C.\ 2012, \araa, 50, 455

\bibitem[Feain et al.(2011)]{2011ApJ...740...17F} Feain, I.~J., Cornwell, T.~J., Ekers, R.~D., et al.\ 2011, \apj, 740, 17

 \bibitem[Gaspari et al.(2017)]{2017MNRAS.466..677G} Gaspari, M., Temi, P., \& Brighenti, F.\ 2017, \mnras, 466, 677


\bibitem[Gendron-Marsolais et al.(2021)]{2021ApJ...911...56G} Gendron-Marsolais, M.-L., Hull, C.~L.~H., Perley, R., et al.\ 2021, \apj, 911, 56

\bibitem[Greisen(2003)]{2003ASSL..285..109G} Greisen, E.~W.\ 2003, Information Handling in Astronomy - Historical Vistas, 285, 109. doi:10.1007/0-306-48080-8\_7

\bibitem[Giacintucci et al.(2007)]{2007A&A...476...99G} Giacintucci, S., Venturi, T., Murgia, M., et al.\ 2007, \aap, 476, 99

\bibitem[Giacintucci et al.(2020)]{2020ApJ...891....1G} Giacintucci, S., Markevitch, M., Johnston-Hollitt, M., et al.\ 2020, \apj, 891, 1 (G20)

\bibitem[Giacintucci et al.(2022)]{2022ApJ...934...49G} Giacintucci, S., Venturi, T., Markevitch, M., et al.\ 2022, \apj, 934, 49. doi:10.3847/1538-4357/ac7805

 
\bibitem[Gizani \& Leahy(2003)]{2003MNRAS.342..399G} Gizani, N.~A.~B. \& Leahy, J.~P.\ 2003, \mnras, 342, 399. doi:10.1046/j.1365-8711.2003.06469.x



\bibitem[Govoni et al.(2005)]{2005A&A...430L...5G} Govoni, F., Murgia, M., Feretti, L., et al.\ 2005, \aap, 430, L5. doi:10.1051/0004-6361:200400113

 \bibitem[Govoni et al.(2009)]{2009A&A...499..371G} Govoni, F., Murgia, M., Markevitch, M., et al.\ 2009, \aap, 499, 371 

\bibitem[Hardcastle(2013)]{2013MNRAS.433.3364H} Hardcastle, M.~J.\ 2013, \mnras, 433, 3364

\bibitem[Hardcastle et al.(2019)]{2019MNRAS.488.3416H} Hardcastle, M.~J., Croston, J.~H., Shimwell, T.~W., et al.\ 2019, \mnras, 488, 3416

\bibitem[Harwood(2017)]{2017MNRAS.466.2888H} Harwood, J.~J.\ 2017, \mnras, 466, 2888. doi:10.1093/mnras/stw3318

\bibitem[Hlavacek-Larrondo et al.(2022)]{2022hxga.book....5H} Hlavacek-Larrondo, J., Li, Y., \& Churazov, E.\ 2022, Handbook of X-ray and Gamma-ray Astrophysics. Edited by Cosimo Bambi and Andrea Santangelo, 5 

\bibitem[Hodgson et al.(2021)]{2021ApJ...909..198H} Hodgson, T., Bartalucci, I., Johnston-Hollitt, M., et al.\ 2021, \apj, 909, 198

\bibitem[Hurley-Walker et al.(2017)]{2017MNRAS.464.1146H} Hurley-Walker, N., Callingham, J.~R., Hancock, P.~J., et al.\ 2017, \mnras, 464, 1146

\bibitem[Ignesti et al.(2020)]{2020A&A...643A.172I} Ignesti, A., Shimwell, T., Brunetti, G., et al.\ 2020, \aap, 643, A172. doi:10.1051/0004-6361/202039056

\bibitem[Intema et al.(2009)]{2009A&A...501.1185I} Intema, H.~T., van der Tol, S., Cotton, W.~D., et al.\ 2009, \aap, 501, 1185. doi:10.1051/0004-6361/200811094

\bibitem[Intema et al.(2017)]{2017A&A...598A..78I} Intema, H.~T., Jagannathan, P., Mooley, K.~P., et al.\ 2017, \aap, 598, A78

\bibitem[Jaffe \& Perola(1973)]{1973A&A....26..423J} Jaffe, W.~J. \& Perola, G.~C.\ 1973, \aap, 26, 423

\bibitem[Jamrozy et al.(2004)]{2004A&A...427...79J} Jamrozy, M., Klein, U., Mack, K.-H., et al.\ 2004, \aap, 427, 79

\bibitem[Jones \& Forman(1984)]{1984ApJ...276...38J} Jones, C. \& Forman, W.\ 1984, \apj, 276, 38

\bibitem[Jun et al.(1995)]{1995ApJ...453..332J} Jun, B.-I., Norman, M.~L., \& Stone, J.~M.\ 1995, \apj, 453, 332


 \bibitem[Katz-Stone, \& Rudnick(1997)]{1997ApJ...488..146K} Katz-Stone, D.~M., \& Rudnick, L.\ 1997, \apj, 488, 146



\bibitem[Knowles et al.(2022)]{2022A&A...657A..56K} Knowles, K., Cotton, W.~D., Rudnick, L., et al.\ 2022, \aap, 657, A56. doi:10.1051/0004-6361/202141488

\bibitem[Koribalski et al.(2024)]{kori24} Koribalski, B.~S., Duchesne, S.~W., Lenc, E., et al.\ 2024, \mnras, ASKAP reveals the radio tail structure of the Corkscrew Galaxy shaped by its passage through the Abell 3627 cluster, 533, 1, 608. doi:10.1093/mnras/stae1838


\bibitem[Laing et al.(2011)]{2011MNRAS.417.2789L} Laing, R.~A., Guidetti, D., Bridle, A.~H., et al.\ 2011, \mnras, 417, 2789


\bibitem[Lal et al.(2022)]{2022ApJ...934..170L} Lal, D.~V., Lyskova, N., Zhang, C., et al.\ 2022, \apj, 934, 170. doi:10.3847/1538-4357/ac7a9b


\bibitem[Lane et al.(2014)]{2014MNRAS.440..327L} Lane, W.~M., Cotton, W.~D., van Velzen, S., et al.\ 2014, \mnras, 440, 327. doi:10.1093/mnras/stu256



\bibitem[Maccagni et al.(2020)]{2020A&A...634A...9M} Maccagni, F.~M., Murgia, M., Serra, P., et al.\ 2020, \aap, 634, A9

\bibitem[Markevitch et al.(2025)]{markevitch25} Markevitch, M., Wik, D., Giacintucci, S., in preparation

\bibitem[Mauch et al.(2003)]{2003MNRAS.342.1117M} Mauch, T., Murphy, T., Buttery, H.~J., et al.\ 2003, \mnras, 342, 1117. doi:10.1046/j.1365-8711.2003.06605.x

 \bibitem[McNamara et al.(2005)]{2005Natur.433...45M} McNamara, B.~R., Nulsen, P.~E.~J., Wise, M.~W., et al.\ 2005, \nat, 433, 45 

 \bibitem[McNamara, \& Nulsen(2012)]{2012NJPh...14e5023M} McNamara, B.~R., \& Nulsen, P.~E.~J.\ 2012, New Journal of Physics, 14, 055023


 \bibitem[Murgia et al.(2010)]{2010A&A...514A..76M} Murgia, M., Eckert, D., Govoni, F., et al.\ 2010, \aap, 514, A76  

 \bibitem[Murgia et al.(2011)]{2011A&A...526A.148M} Murgia, M., Parma, P., Mack, K.-H., et al.\ 2011, \aap, 526, A148


\bibitem[Murphy et al.(2007)]{2007MNRAS.382..382M} Murphy, T., Mauch, T., Green, A., et al.\ 2007, \mnras, 382, 382. doi:10.1111/j.1365-2966.2007.12379.x


\bibitem[Nolting et al.(2019a)]{2019ApJ...876..154N} Nolting, C., Jones, T.~W., O'Neill, B.~J., et al.\ 2019, \apj, 876, 154. doi:10.3847/1538-4357/ab16d6

\bibitem[Nolting et al.(2019b)]{2019ApJ...885...80N} Nolting, C., Jones, T.~W., O'Neill, B.~J., et al.\ 2019, \apj, 885, 80. doi:10.3847/1538-4357/ab4650

 \bibitem[Nulsen et al.(2005)]{2005ApJ...628..629N} Nulsen, P.~E.~J., McNamara, B.~R., Wise, M.~W., \& David, L.~P.\ 2005, \apj, 628, 629 

\bibitem[Offringa et al.(2014)]{2014MNRAS.444..606O} Offringa, A.~R., McKinley, B., Hurley-Walker, N., et al.\ 2014, \mnras, 444, 606

\bibitem[Offringa \& Smirnov(2017)]{2017MNRAS.471..301O} Offringa, A.~R. \& Smirnov, O.\ 2017, \mnras, 471, 301. doi:10.1093/mnras/stx1547


\bibitem[Owen et al.(2014)]{2014ApJ...794...24O} Owen, F.~N., Rudnick, L., Eilek, J., et al.\ 2014, \apj, 794, 24. doi:10.1088/0004-637X/794/1/24

\bibitem[Owen et al.(2000)]{2000ApJ...543..611O} Owen, F.~N., Eilek, J.~A., \& Kassim, N.~E.\ 2000, \apj, 543, 611. doi:10.1086/317151


\bibitem[Pandge et al.(2022)]{2022MNRAS.509.1837P} Pandge, M.~B., Kale, R., Dabhade, P., et al.\ 2022, \mnras, 509, 1837. doi:10.1093/mnras/stab2945


\bibitem[Parma et al.(2007)]{2007A&A...470..875P} Parma, P., Murgia, M., de Ruiter, H.~R., et al.\ 2007, \aap, 470, 875



\bibitem[Parrish et al.(2012)]{2012MNRAS.422..704P} Parrish, I.~J., McCourt, M., Quataert, E., et al.\ 2012, \mnras, 422, 704. doi:10.1111/j.1365-2966.2012.20650.x

\bibitem[Paterno-Mahler et al.(2013)]{2013ApJ...773..114P} Paterno-Mahler, R., Blanton, E.~L., Randall, S.~W., et al.\ 2013, \apj, 773, 2, 114. doi:10.1088/0004-637X/773/2/114


\bibitem[Peres et al.(1998)]{1998MNRAS.298..416P} Peres, C.~B., Fabian, A.~C., Edge, A.~C., et al.\ 1998, \mnras, 298, 416

\bibitem[Perley et al.(1984)]{1984ApJ...285L..35P} Perley, R.~A., Dreher, J.~W., \& Cowan, J.~J.\ 1984, \apjl, 285, L35

 \bibitem[Perley, \& Butler(2017)]{2017ApJS..230....7P} Perley, R.~A., \& Butler, B.~J.\ 2017, \apjs, 230, 7





\bibitem[Peterson \& Fabian(2006)]{2006PhR...427....1P} Peterson, J.~R. \& Fabian, A.~C.\ 2006, \physrep, 427, 1

\bibitem[Pulido et al.(2018)]{2018ApJ...853..177P} Pulido, F.~A., McNamara, B.~R., Edge, A.~C., et al.\ 2018, \apj, 853, 177

\bibitem[Rajpurohit et al.(2020)]{2020A&A...636A..30R} Rajpurohit, K., Hoeft, M., Vazza, F., et al.\ 2020, \aap, 636, A30. doi:10.1051/0004-6361/201937139


\bibitem[Rajpurohit et al.(2024)]{2024ApJ...976...64R} Rajpurohit, K., O'Sullivan, E., Schellenberger, G., et al.\ 2024, \apj, 976, 64. doi:10.3847/1538-4357/ad8136



\bibitem[Ramatsoku et al.(2020)]{2020A&A...636L...1R} Ramatsoku, M., Murgia, M., Vacca, V., et al.\ 2020, \aap, 636, L1. doi:10.1051/0004-6361/202037800

\bibitem[Randriamanakoto et al.(2020)]{2020MNRAS.496.3381R} Randriamanakoto, Z., Ishwara-Chandra, C.~H., \& Taylor, A.~R.\ 2020, \mnras, 496, 3381. doi:10.1093/mnras/staa1782


\bibitem[Rengelink et al.(1997)]{1997A&AS..124..259R} Rengelink, R.~B., Tang, Y., de Bruyn, A.~G., et al.\ 1997, \aaps, 124, 259. doi:10.1051/aas:1997358


\bibitem[Rudnick(2002)]{2002NewAR..46...95R} Rudnick, L.\ 2002, \nar, 46, 95. doi:10.1016/S1387-6473(01)00162-2


\bibitem[Rudnick et al.(2021)]{2021Galax...9...81R} Rudnick, L., Cotton, W., Knowles, K., et al.\ 2021, Galaxies, 9, 81. doi:10.3390/galaxies9040081


\bibitem[Rudnick et al.(2022)]{2022ApJ...935..168R} Rudnick, L., Br{\"u}ggen, M., Brunetti, G., et al.\ 2022, \apj, 935, 168

 \bibitem[Scaife, \& Heald(2012)]{2012MNRAS.423L..30S} Scaife, A.~M.~M., \& Heald, G.~H.\ 2012, \mnras, 423, L30


\bibitem[Schekochihin et al.(2004)]{2004ApJ...612..276S} Schekochihin, A.~A., Cowley, S.~C., Taylor, S.~F., et al.\ 2004, \apj, 612, 276


\bibitem[Schlafly et al.(2018)]{2018ApJS..234...39S} Schlafly, E.~F., Green, G.~M., Lang, D., et al.\ 2018, \apjs, The DECam Plane Survey: Optical Photometry of Two Billion Objects in the Southern Galactic Plane, 234, 2, 39

\bibitem[Shimwell et al.(2016)]{2016MNRAS.459..277S} Shimwell, T.~W., Luckin, J., Br{\"u}ggen, M., et al.\ 2016, \mnras, 459, 277


\bibitem[Shulevski et al.(2017)]{2017A&A...600A..65S} Shulevski, A., Morganti, R., Harwood, J.~J., et al.\ 2017, \aap, 600, A65


\bibitem[Shulevski et al.(2024)]{2024A&A...682A.171S} Shulevski, A., Brienza, M., Massaro, F., et al.\ 2024, \aap, 682, A171


\bibitem[Slee et al.(2001)]{2001AJ....122.1172S} Slee, O.~B., Roy, A.~L., Murgia, M., et al.\ 2001, \aj, 122, 1172

 

\bibitem[Timmerman et al.(2022)]{2022A&A...658A...5T} Timmerman, R., van Weeren, R.~J., Callingham, J.~R., et al.\ 2022, \aap, 658, A5


 \bibitem[Tregillis et al.(2004)]{2004ApJ...601..778T} Tregillis, I.~L., Jones, T.~W., \& Ryu, D.\ 2004, \apj, 601, 778

\bibitem[Upreti et al.(2024)]{upreti24} Upreti, N., Vaidya, B., \& Shukla, A.\ 2024, Journal of High Energy Astrophysics, Bridging simulations of kink instability in relativistic magnetized jets with radio emission and polarisation, 44, 146. doi:10.1016/j.jheap.2024.09.007



\bibitem[Vantyghem et al.(2019)]{2019ApJ...870...57V} Vantyghem, A.~N., McNamara, B.~R., Russell, H.~R., et al.\ 2019, \apj, 870, 57

\bibitem[van Weeren et al.(2024)]{2024A&A...692A..12V} van Weeren, R.~J., Timmerman, R., Vaidya, V., et al.\ 2024, \aap, 692, A12. doi:10.1051/0004-6361/202451618
 
\bibitem[Vazza et al.(2021)]{2021A&A...653A..23V} Vazza, F., Wittor, D., Brunetti, G., et al.\ 2021, \aap, 653, A23

\bibitem[Venturi et al.(2017)]{venturi17} Venturi, T., Rossetti, M., Brunetti, G., et al.\ 2017, \aap, 603, A125


\bibitem[Venturi et al.(2022)]{2022A&A...660A..81V} Venturi, T., Giacintucci, S., Merluzzi, P., et al.\ 2022, \aap, 660, A81

 \bibitem[Vikhlinin et al.(2005)]{2005ApJ...628..655V} Vikhlinin, A., Markevitch, M., Murray, S.~S., et al.\ 2005, \apj, 628, 655


 \bibitem[Voit et al.(2015)]{2015Natur.519..203V} Voit, G.~M., Donahue, M., Bryan, G.~L., et al.\ 2015, \nat, 519, 203



 \bibitem[Werner et al.(2016)]{2016MNRAS.460.2752W} Werner, N., Zhuravleva, I., Canning, R.~E.~A., et al.\ 2016, \mnras, 460, 2752  

 \bibitem[Wise et al.(2007)]{2007ApJ...659.1153W} Wise, M.~W., McNamara, B.~R., Nulsen, P.~E.~J., et al.\ 2007, \apj, 659, 1153

\bibitem[Wykes et al.(2013)]{2013A&A...558A..19W} Wykes, S., Croston, J.~H., Hardcastle, M.~J., et al.\ 2013, \aap, 558, A19

\bibitem[Wykes et al.(2014)]{2014MNRAS.442.2867W} Wykes, S., Intema, H.~T., Hardcastle, M.~J., et al.\ 2014, \mnras, 442, 286

\bibitem[ZuHone et al.(2013)]{zuhone13} ZuHone, J.~A., Markevitch, M., Brunetti, G., et al.\ 2013, \apj, 762, 2, 78

\bibitem[ZuHone et al.(2021)]{2021ApJ...914...73Z} ZuHone, J.~A., Markevitch, M., Weinberger, R., et al.\ 2021, \apj, 914, 73


\end{thebibliography}
\end{document}